\documentclass[prb,reprint,showpacs,superscriptaddress]{revtex4-1}
\usepackage{graphicx}
\usepackage{amsfonts}
\usepackage[figuresright]{rotating}
\usepackage{float}
\usepackage{amssymb}
\usepackage{amsmath}
\usepackage{hyperref}

\def\avg#1{\langle#1\rangle}

\def\be{\begin{equation}} \def\ee{\end{equation}}
\def\bea{\begin{eqnarray}} \def\eea{\end{eqnarray}}

\def\nn{\nonumber}
\def\pp{\parallel}

\begin{document}
\title{The honeycomb lattice with multi-orbital structure:
topological and quantum anomalous Hall insulators with large
gaps}
\author{Gu-Feng Zhang}
\email{guz003@physics.ucsd.edu}
\affiliation{Department of Physics, University of California, San Diego,
CA 92093}
\author{Yi Li}
\email{YL5@princeton.edu}
\affiliation{Princeton Center for Theoretical Science, Princeton University,
Princeton, NJ 08544}
\author{Congjun Wu}
\affiliation{Department of Physics, University of California, San Diego,
CA 92093}
\begin{abstract}
  We construct a minimal four-band model for the two-dimensional (2D) topological insulators and quantum anomalous Hall
  insulators based on the $p_x$- and $p_y$-orbital bands in the honeycomb lattice.  The multiorbital structure allows
  the atomic spin-orbit coupling which lifts the degeneracy between two sets of on-site Kramers doublets
  $j_z=\pm\frac{3}{2}$ and $j_z=\pm\frac{1}{2}$.  Because of the orbital angular momentum structure of Bloch-wave states
  at $\Gamma$ and $K(K^\prime)$ points, topological gaps are equal to the atomic spin-orbit coupling strengths, which
  are much larger than those based on the mechanism of the $s$-$p$ band inversion.  In the weak and intermediate regime of
  spin-orbit coupling strength, topological gaps are the global gap.  The energy spectra and eigen wave functions are
  solved analytically based on Clifford algebra.  The competition among spin-orbit coupling $\lambda$, sublattice
  asymmetry $m$ and the N\'eel exchange field $n$ results in band crossings at $\Gamma$ and $K (K^\prime)$ points, which
  leads to various topological band structure transitions. The quantum anomalous Hall state is reached under the
  condition that three gap parameters $\lambda$, $m$, and $n$ satisfy the triangle inequality.  Flat bands also
  naturally arise which allow a local construction of eigenstates.  The above mechanism is related to several classes of
  solid state semiconducting materials.
\end{abstract}
\maketitle

\section{Introduction}
The two-dimensional (2D) quantum Hall effect \cite{klitzing1980} is among the early examples of topological states of
matter whose magnetic band structure is characterized by the first Chern number
\cite{thouless1982,halperin1982,kohmoto1985,haldane1988}.  Later on, quantum anomalous Hall (QAH) insulators were
proposed with Bloch band structures \cite{haldane1988}.  Insulators with nontrivial band topology were also generalized
into time-reversal (TR) invariant systems, termed topological insulators (TIs) in both 2D and 3D, which have become a
major research focus in contemporary condensed matter physics \cite{qi2010a,qi2011,Hasan2010}.  The topological index of
TR invariant TIs is no longer just integer valued, but $\mathbb{Z}_2$ valued, in both 2D and 3D
\cite{kane2005,bernevig2006,bernevig2006a, qi2008a,fu2007a,moore2007,roy2010}.  In 4D, it is the integer-valued second
Chern number \cite{zhang2001,qi2008a}.  Various 2D and 3D TI materials were predicted theoretically and observed
experimentally \cite{bernevig2006a,konig2007,hsieh2008,zhang2009b,xia2009,chen2009a}.  They exhibit gapless helical 1D
edge modes and 2D surface modes through transport and spectroscopic measurements.

Solid state materials with the honeycomb lattice structure (e.g., graphene) are another important topic of condensed
matter physics \cite{neto2009,novoselov2005,zhang2005}. There are several proposals of QAH model in the honeycomb
lattice\cite{Zhang2013, Wu2014}.  As a TR invariant doublet of Haldane's QAH model \cite{kane2005,kane2005a}, the
celebrated 2D Kane-Mele model was originally proposed in the context of graphene-like systems with the $p_z$ band.
However, the atomic level spin-orbit (SO) coupling in graphene does not directly contribute to opening the topological
band gap \cite{min2006}. Because of the single band structure and the lattice symmetry, the band structure SO coupling
is at the level of a high-order perturbation theory and thus is tiny.

Recently, the $p_x$- and $p_y$-orbital physics in the honeycomb lattice has been systematically investigated in the
context of ultracold-atom optical lattices \cite{wu2007,wu2008b,wu2008a,wu2008, zhang2010,lee2010,zhang2011a}.  The
optical potential around each lattice potential minimum is locally harmonic.  The $s$- and $p$-orbital bands are
separated by a large band gap, and thus the hybridization between them is very small.  The $p_z$-orbital band can also
be tuned to high energy by imposing strong laser beams along the $z$ direction.  Consequently, we can have an ideal
$p_x$- and $p_y$-orbital system in the artificial honeycomb optical lattice.

Such an orbitally active system provides a great opportunity to investigate the interplay between nontrivial band
topology and strong correlations, which is fundamentally different from graphene \cite{wu2007,wu2008a,wu2008}.  Its band
structure includes not only Dirac cones but also two additional narrow bands which are exactly flat in the limit of
vanishing $\pi$ bonding.  Inside the flat bands, due to the vanishing kinetic energy scale, nonperturbative strong
correlation effects appear, such as the Wigner crystallization of spinless fermions \cite{wu2007,wu2008a} and
ferromagnetism \cite{zhang2010} of spinful fermions as {\it exact} solutions. Very recently, the honeycomb lattice for
polaritons has been fabricated \cite{Jacqmin2014}.  Both the Dirac cone and the flat dispersion for the $p_x/p_y$
orbital bands have been experimentally observed. The band structure can be further rendered topologically nontrivial by
utilizing the existing experimental technique of the on-site rotation around each trap center
\cite{gemelke2010}.  This provides a natural way to realize the QAH effect (QAHE) as proposed in
Refs. \onlinecite{wu2008} and \onlinecite{zhang2011a}, and the topological gaps are just the rotation angular velocity
\cite{wu2008,zhang2011a}.  In the Mott-insulating states, the frustrated orbital exchange can be described by a novel
quantum 120$^\circ$ model \cite{wu2008b}, whose classic ground states map to all the possible loop configurations in the
honeycomb lattice.  The $p_x$- and $p_y$-orbital structure also enables unconventional $f$-wave Cooper pairing even with
conventional interactions exhibiting flat bands of zero energy Majorana edge modes along boundaries parallel to gap
nodal directions \cite{lee2010}.

The $p_x$- and $p_y$-orbital structures have also been studied very
recently in several classes of solid state semiconducting materials
including fluoridated tin film
\cite{xu2013,wang2014,Wu2014}, functionalized germanene systems
\cite{si2014a}, Bi$X$/Sb$X$ ($X$=H,F,Cl,Br) systems
\cite{liu2014,song2014},
and in organic materials \cite{wang2013a,wang2013,liu2013}.
All these materials share the common feature of the
active $p_x$ and $p_y$ orbitals in the honeycomb lattice,
enabling a variety of rich structures of topological band physics.
The most striking property is the prediction of the large
topological band gap which can even exceed room temperature.

In the literature, a common mechanism giving rise to topological band gaps is the band inversion, which typically
applies for two bands with different orbital characters, say, the $s$-$p$ bands. However, although band inversion
typically occurs in systems with strong SO coupling, the SO coupling does not directly contribute to the value of the
gap. The band inversion would lead to gap closing at finite momenta in the absence of the $s$-$p$ hybridization, and the
$s$-$p$ hybridization reopens the gap whose nature becomes topological. The strength of the hybridization around the
$\Gamma$ point linearly depends on the magnitude of the momenta, in the spirit of the $k\cdot p$ perturbation theory,
which is typically small. This is why in usual topological insulators based on band inversion, in spite of considerable
SO coupling strengths, the topological gap values are typically small. On the other hand, as for the single band systems
in the honeycomb lattice such as graphene, the effect from the atomic level SO coupling to the band structure is also
tiny, as a result of the high-order perturbation theory.

In the model presented in this paper, here are only $p$ orbitals. The two-sublattice structure and the $p_x/p_y$-orbital
configuration together greatly enhance the effect of SO coupling, as illustrated in Fig.~\ref{fig:gap}. The atomic-scale
SO coupling directly contributes to the opening of the topological gap at the $K(K')$ point between bands 2 and 3, and
that at the $\Gamma$ point between bands 1 and 2. Since the atomic SO coupling can be very large, the topological band
gap can even reach the level of $0.3\,eV$ according to the estimation in Ref.\,\onlinecite{si2014a}.

In this article, we construct a minimal four-band model to analyze the topological properties based on the $p_x$- and
$p_y$-orbital structure in the honeycomb lattice.  The eigen energy spectra and wave functions can be analytically
solved with the help of Clifford $\Gamma$ matrices.  The atomic SO coupling lifts the degeneracy between two on-site
Kramers pairs with $j_z=\pm\frac{3}{2}$ and $j_z=\pm\frac{1}{2}$.  As explained in the preceding paragraph, the
topological gap in this class of systems is extraordinary large.  In the weak and intermediate regime of spin-orbit
coupling strength, the topological gaps are the global gap.  The lattice asymmetry and the SO coupling provide two
different gap opening mechanisms, and their competition leads to a variety of topological band structures.  With the
introduction of both the sublattice anisotropy and the N\'eel exchange field, the system can become a large gap QAH
insulator.

The article is organized as follows.  The four-band model for the $p_x$- and $p_y$-orbital system in the honeycomb
lattice is constructed in Sec.~\ref{sect:model}.  The symmetry analysis is presented in Sec.~\ref{sect:symm}.  In
Sec.~\ref{sect:solution}, the analytic solutions of energy spectra and eigen wave functions are presented.  The study of
band topology and band crossing is presented in Sec.~\ref{sect:topo}.  Effective two-band models are constructed around
high-symmetry points near band crossings in Sec.~\ref{sect:twoband}.  The mechanism of large topological band gap is
explained in Sec.~\ref{sect:largegap}.  We add the N\'eel exchange field term in Sec.~\ref{sect:QAHE}, and investigate
how to get a large gap QAH insulator.  Conclusions are presented in Sec.~\ref{sect:conclusion}.


\section{The $p_x$ and $p_y$ band Hamiltonian}
\label{sect:model}
The two sublattices of the honeycomb lattice are denoted $A$
and $B$.
The bonding part of the Hamiltonian is
\bea
H_0&=&t_{\pp} \sum_{ \vec{r} \in A,s} \big\{ p^\dagger_{i,s}
(\vec r) p_{i,s}(\vec r +a \hat e_i)+{\rm H.c.} \big\} \nn \\
&-&t_{\perp} \sum_{ \vec{r} \in A,s} \big\{ p^{\prime\dagger}_{i,s}(\vec r)
p^\prime_{i,s}(\vec{r}+a\hat{e}_i) + {\rm H.c.} \big\},
\label{eq:ham0}
\eea where $s=\uparrow,\downarrow$ represents two eigenstates of spin $s_z$;
$\hat{e}_{1,2}=\pm\frac{\sqrt{3}}{2}\hat{e}_x+\frac{1}{2}\hat{e}_y$ and $\hat{e}_3=-\hat{e}_y$ are three unit vectors
from one $A$ site to its three neighboring $B$ sites; $a$ is the nearest neighbor bond length; $p_i\equiv
(p_x\hat{e}_x+p_y\hat{e}_y)\cdot \hat{e}_i$ and $p^\prime_i\equiv(-p_x\hat{e}_y+p_y\hat{e}_x)\cdot \hat{e}_i$ are the
projections of the $p$ orbitals parallel and perpendicular to the bond direction $\hat e_i$ for $i=1,\cdots,3$,
respectively; $t_\parallel$ and $t_\perp$ are the corresponding $\sigma$- and $\pi$-bonding strengths, respectively.
Typically speaking, $t_\perp$ is much smaller than $t_\pp$.  The signs of the $\sigma$- and $\pi$-bonding terms are
opposite to each other because of the odd parity of $p$-orbitals.  The $p_z$ orbital is inactive because it forms
$\sigma$ bonding with halogen atoms or the hydrogen atom.

There exists the atomic SO coupling $\vec s \cdot \vec L$ on each site.  However, under the projection into the $p_x$-
and $p_y$-orbital states, there are only four on-site single-particle states.  They can be classified into two sets of
Kramers doublets: $p^\dagger_{+,\uparrow}|0\rangle$ and $p_{-,\downarrow}^\dagger|0\rangle$ with $j_z=\pm\frac{3}{2}$,
and $p_{+,\downarrow}^\dagger|0\rangle$ and $p_{-,\uparrow}^\dagger|0\rangle$ with $j_z=\pm\frac{1}{2}$, where
$p^\dagger_{\pm,s}=\frac{1}{\sqrt 2} (p^\dagger_{x,s}\pm i p^\dagger_{y,s})$ are the orbital angular momentum $L_z$
eigenstates and $j_z$ is the $z$ component of total angular momentum.  These four states cannot be mixed under $j_z$
conservation, and thus only the $s_z L_z$ term survives which splits the degeneracy between the two sets of Kramers
doublets.  The SO coupling is modeled as \bea H_{so}=-\lambda\sum_{\vec r,\sigma,s} \sigma\,s \,
p^\dagger_{\sigma,s}(\vec r)p_{\sigma,s} (\vec r),
\label{eq:so}
\eea
where $\sigma=\pm$ refers to the orbital angular momentum number
$L_z$, $s=\pm$ corresponds to the eigenvalues of $s_z=
\uparrow,\downarrow$, and $\lambda$ is the SO coupling strength.
For completeness, we also add the sublattice asymmetry term
\bea
H_m&=&m\Big\{\sum_{\vec r \in A,\sigma,s}
p^\dagger_{\sigma,s}(\vec r) p_{\sigma,s} (\vec r)
-\sum_{\vec r \in B,\sigma,s}
p^\dagger_{\sigma,s}(\vec r) p_{\sigma,s} (\vec r)
\Big\}.\nn \\
\label{eq:m}
\eea

In Sec.~\ref{sect:QAHE}, we will consider the QAH state based on this system by adding the following time-reversal (TR)
symmetry breaking N\'eel exchange term
\bea
H_n&=&n\Big\{\sum_{\vec r \in A,\sigma,s}
s\,p^\dagger_{\sigma,s}(\vec r) p_{\sigma,s} (\vec r) \nn \\
&-&\sum_{\vec r \in B,\sigma,s} s\,p^\dagger_{\sigma,s}(\vec r) p_{\sigma,s} (\vec r) \Big\}.
\label{eq:neel}
\eea
where $n$ is the N\'eel exchange field strength.
Before Sec.~\ref{sect:QAHE}, we only consider the Hamiltonian
$H_0+H_{so}+H_{m}$ without the N\'eel exchange term.

\section{Symmetry properties}
\label{sect:symm}

One key observation is that electron spin $s_z$ is conserved for the
total Hamiltonian $H_0+H_{so}+H_m$.
We will analyze the band structure in the sector with $s=\uparrow$, and
that with $s=\downarrow$ can be obtained by performing time-reversal
(TR) transformation.
$H_0+H_{so}$ is a TR doubled version of the QAH model proposed in
ultracold fermion systems in honeycomb optical lattices
\cite{wu2008}.
In the sector with $s=\uparrow$, we introduce the four-component spinor
representation in momentum space defined as
\bea
\psi_{\uparrow\tau\sigma}(\vec k)&=&(\psi_{\uparrow,A,+}(\vec k),
\psi_{\uparrow,B,+}(\vec k), \nn \\
&&\psi_{\uparrow,A,-}(\vec k), \psi_{\uparrow,B,-}(\vec k))^T,
\label{eq:bases1}
\eea
where two sublattice components are denoted $A$ and $B$.
The doublet of orbital angular momentum and that
of the sublattice structure are considered as two
independent pseudospin degrees of freedom, which are
denoted by two sets of Pauli matrices as
$\sigma_{1,2,3}$ and $\tau_{1,2,3}$, respectively.
Unlike $s_z$, these two pseudospins are not conserved.
The nearest neighbor hopping connects $A$-$B$ sublattices,
which does not conserve the orbital angular momentum
due to orbital anisotropy in lattice systems.

The Hamiltonian $H_{\uparrow}(\vec k)$ can be conveniently represented as
\bea
H_{\uparrow}(\vec k)&=&h_{03} 1_\tau\otimes \sigma_3
+ h_{30} \tau_3 \otimes 1_\sigma
+h_{10}(\vec k)\tau_1 \otimes 1_\sigma \nn \\
&+&h_{20} (\vec k) \tau_2 \otimes 1_\sigma
+h_{11} (\vec k) \tau_1\otimes \sigma_1 + h_{22} (\vec k)
\tau_2\otimes \sigma_2 \nn \\
&+&h_{21} (\vec k) \tau_2\otimes \sigma_1
+ h_{12} (\vec k) \tau_1\otimes \sigma_2,
\eea
with the expressions of
\bea
h_{03}&=&-\lambda, \ \ \, \ \ \, h_{30}=m, \nn \\
h_{10}&=& t_1\sum_{i=1}^3 \cos (\vec k \cdot \hat e_i), \nn \\
h_{20}&=& -t_1\sum_{i=1}^3 \sin (\vec k \cdot \hat e_i), \nn \\
h_{11}&=& t_2\sum_{i=1}^3 \cos (\vec k \cdot \hat e_i) \cos2\theta_i, \nn \\
h_{22}&=& -t_2\sum_{i=1}^3 \sin (\vec k \cdot \hat e_i) \sin2\theta_i, \nn \\
h_{21}&=& -t_2\sum_{i=1}^3 \sin (\vec k \cdot \hat e_i) \cos2\theta_i, \nn \\
h_{12}&=& t_2\sum_{i=1}^3 \cos(\vec k \cdot \hat e_i) \sin2\theta_i,
\label{eq:HK_up}
\eea
where $t_{1,2}=\frac{1}{2}(t_\pp\pm t_\perp)$ and
$\theta_i=\frac{1}{6}\pi,\frac{5}{6}\pi$, $\frac{3}{2}\pi$
are the azimuthal angles of the bond orientation $\hat e_{i}$
for $i=1,2$ and 3, respectively.

For the sector with $s=\downarrow$, the four-component
spinors $\psi_{\downarrow}$ are constructed as
$\psi_{\downarrow\tau\sigma}(\vec k)=(\psi_{\downarrow,A,+}(\vec k),
\psi_{\downarrow,B,+}(\vec k), \psi_{\downarrow,A,-}(\vec k),
\psi_{\downarrow,B,-}(\vec k))^T$.
Under this basis, $H_{\downarrow}(\vec k)$ has the same
matrix form as that of $H_{\uparrow}(\vec k)$ except we flip
the sign of $\lambda$ in the $h_{03}$ term.

Next we discuss the symmetry properties of $H_{\uparrow}(\vec k)$.
We first consider the case of $m=0$, i.e.,
in the absence of the lattice asymmetry.
$H_{\uparrow}(\vec k)$ satisfies the parity symmetry defined as
\bea
P H_\uparrow (\vec k) P^{-1}= H_\uparrow (-\vec k),
\label{eq:parity}
\eea
with $P=\tau_1\otimes 1_\sigma$.
$H_{\uparrow}(\vec k)$ also possesses the particle-hole 
symmetry
\bea
C^\prime H_\uparrow(\vec k) (C^{\prime})^{-1}=-H_\uparrow^*(-\vec k),
\label{eq:ch_1}
\eea
where $C^\prime=\tau_3\otimes \sigma_1$, satisfying
$(C^{\prime})^2=1$, and $*$ represents complex conjugation.
$C^\prime$ is the operation of $p_{\uparrow, A, \sigma}
\rightarrow p_{\uparrow, A, \sigma}$ and
$p_{\uparrow, B, \sigma} \rightarrow -p_{\uparrow, B, \sigma}$
combined with switching eigenstates of $L_z$.

Furthermore, when combining two sectors of $s=\uparrow$
and $\downarrow$ together, the system satisfies the TR symmetry
defined as  $T=is_2 \otimes 1_\tau \otimes \sigma_1\otimes K$
with $T^2=-1$, where $K$ is the complex conjugation.
Due to the above symmetry proprieties, our system is
in the DIII class \cite{schnyder2008} in the absence of lattice asymmetry.
However, in the presence of lattice asymmetry, the particle-hole 
symmetry $C^\prime$ is broken, and only the TR symmetry exists.
In that case, the system is the in sympletic class AII.
In both cases, the topological index is $\mathbb{Z}_2$.

Nevertheless, in the presence of sublattice asymmetry $m$,
the product of parity  and particle-hole transformations
remains a valid symmetry as
\bea
C H_{\uparrow}(\vec k) C^{-1}= - H^*_\uparrow(\vec k),
\label{eq:ch_2}
\eea
where $C=i\tau_2\otimes \sigma_1$ satisfying $C^2=-1$.
This symmetry ensures the energy levels, for each $\vec k$, appear symmetric with respect to the zero energy.

Without loss of generality, we choose the convention that $m>0$
and $\lambda>0$ throughout the rest of this article.
The case of $m<0$ can be obtained through a parity transformation
that flips the $A$ and $B$ sublattices as
\bea
H_{m<0}(\vec k)= (\tau_1 \otimes 1_\sigma) H_{m>0} (-\vec k)
(\tau_1 \otimes 1_\sigma)^{-1}.
\eea
The case of $\lambda<0$ can be obtained
through a partial
TR transformation only within each spin sector but
without flipping electron spin:
\bea
H_{\lambda<0}(\vec k)= (1_\tau\otimes \sigma_1) H^*_{\lambda>0}(-\vec k)
(1_\tau\otimes \sigma_1)^{-1}.
\eea

\section{Energy spectra and eigenfunctions}
\label{sect:solution}

In this section, we provide solutions to the Hamiltonian
of $p_x$- and $p_y$-orbital bands in honeycomb lattices.
Based on the properties of $\Gamma$ matrices, most results
can be expressed analytically.

\subsection{Analytic solution to eigen energies}
Due to 
Eq. (\ref{eq:ch_2}), the spectra of
$H_\uparrow(\vec k)$ are symmetric with respect to the zero energy.
Consequently, they can be analytically solved as follows.
The square of $H_\uparrow(\vec k)$ can be represented in the
standard $\Gamma$-matrix representation as
\bea
H^2(\vec k)= g_0 (\vec k)+ 2\sum_{i=1}^5 g_i(\vec k) \Gamma_i,
\eea
with the $g_i$'s expressed as
\begin{widetext}
\bea
g_0&=&\lambda^2+ m^2 + 3(t_1^2 +t_2^2)+(2t_2^2-t_1^2)
\sum_{j=1}^3 \cos \vec k \cdot \vec b_j, \nn \\
g_1&=&-t_1t_2 \sum_{j=1}^3 \cos \vec k \cdot \vec b_j \sin \theta_i, \ \ \,
g_5= -t_1t_2 \sum_{j=1}^3 \cos \vec k \cdot \vec b_j \cos\theta_i,
\nn \\
g_2 &=& -\lambda t_2 \sum_{j=1}^3 \cos \vec k \cdot \vec a_j, \ \ \, \ \ \,
g_3= -\lambda t_2 \sum_{j=1}^3 \sin \vec k \cdot \vec a_j,  \nn \\
g_4&=&\frac{\sqrt 3}{2}t_1^2 \sum_{j=1}^3 \sin \vec k \cdot \vec b_j
-m\lambda, \ \ \, \ \ \,
\eea
\end{widetext}
where $\vec b_1=\hat e_2-\hat e_3$,
$\vec b_2=\hat e_3-\hat e_1$, and $\vec b_3=\hat e_1-\hat e_2$.

The $\Gamma$ matrices satisfy the anticommutation relation
as $\{\Gamma_i, \Gamma_j\}=2\delta_{ij}$.
They are defined here as
\bea
\Gamma_1=1_\tau\otimes \sigma_1, \ \ \,
\Gamma_{2,3,4}=\tau_{1,2,3}\otimes \sigma_3, \ \ \,
\Gamma_{5}=1_\tau\otimes \sigma_2. \ \ \
\eea
The spectra are solved as
$E^2(\vec k)=g_0\pm 2 (\sum_{i=1}^5 g_i^2)^{\frac{1}{2}}$.

In the case of neglecting the $\pi$ bonding, i.e., $t_1=t_2=
\frac{1}{2}t_\pp$,
the spectra can be expressed as
\bea
E_{1,4}(\vec k)&=& \pm \sqrt{f_1(\vec k)+\sqrt{f_2(\vec k)} },\nn \\
E_{2,3}(\vec k)&=& \pm \sqrt{ f_1(\vec k)-\sqrt{f_2(\vec k)} },
\label{eq:spectra}
\eea
where
\bea
f_1(\vec k) &=&\lambda^2+m^2+\frac{3}{2}t_\pp^2 +\frac{1}{4}t_\pp^2
\eta_c(\vec k), \nn \\
f_2(\vec k) &=&
\big\{\frac{t^2_\pp}{4}[3-\eta_c(\vec k) ]
-4\lambda^2  \big\}^2 \nn \\
&+& \lambda^2(9t_\pp^2  -16 \lambda^2
+4m^2)-\frac{\sqrt 3}{4}t_\pp^2
m\lambda \eta_s(\vec k), \ \ \
\eea
and the expressions for $\eta_c$, $\eta_s$ are defined as
\bea
\eta_c(\vec k)&=& \sum_{j=1}^3 \cos \vec k \cdot \vec b_j,  \ \ \,
\eta_s(\vec k)=\sum_{j=1}^3 \sin \vec k \cdot \vec b_j.
\eea

\subsection{Solution to eigen wave functions}

Eigen-wave functions $\psi_i(\vec k)$ for the band
index $i=1,\cdots, 4$ can be obtained by applying two steps
of projection operators successively.
The first projection is based on $H^2(\vec k)$ which
separates the subspace spanned by $\psi_{1,4}(\vec k)$
from that by $\psi_{2,3}(\vec k)$.
We define
\bea
P_{14}(\vec k)&=&\frac{1}{2} \big[1+\sum_{i=1}^5
g^\prime_i(\vec k) \Gamma_i\big],\nn \\
P_{23}(\vec k)&=&\frac{1}{2} \big[1- \sum_{i=1}^5
g^\prime_i(\vec k) \Gamma_i\big],
\eea
where $g^\prime_i$ is normalized according to $g_i^\prime(\vec k)=
g_i(\vec k)/\sqrt{f_2(\vec k)}$ such that $\sum_i g^{\prime,2}_i=1$.
In each subspace, we can further distinguish the positive
and negative energy states by applying
\bea
P_i(\vec k) =\frac{1}{2}\big\{ 1+ \frac{1}{E_i}H_\uparrow(\vec k)\big\}.
\eea
for each band $i=1,\cdots, 4$.
In other words, starting from an arbitrary state vector $\psi(\vec k)$,
we can decompose it into $\psi(\vec k)= \sum_{i=1}^4 \phi_i(\vec k)$
according to
\bea
\phi_{1,4}(\vec k)&=&P_{1,4} (\vec k) P_{14}(\vec k)\psi, \nn \\
\phi_{2,3}(\vec k)&=&P_{2,3} (\vec k) P_{23}(\vec k)\psi.
\eea
which satisfy $H\phi_i(\vec k)=E_i \phi_i(\vec k)$.
Nevertheless, the concrete expressions of eigen wave functions
$\psi_i (i=1,\cdots, 4)$ after normalization are rather complicated
and thus we will not present their detailed forms.

\subsection{A new set of bases}
Below we present a simplified case in the absence of SO coupling,
i.e., $\lambda=0$, in which
the two-step diagonalizations can be constructed explicitly.
This also serves as a set of convenient bases for further studying
the band topology after turning on SO coupling.
We introduce a new set of orthonormal bases denoted as
\bea
|A_1(\vec k)\rangle
&=&\frac{1}{\sqrt{2N_k}}
\left(
\begin{array}{c}
\gamma^*_{1-}(\vec k)\\
0\\
\gamma^*_{1+}(\vec k) \\
0\\
\end{array}
\right), \nn \\
|B_1(\vec k)\rangle&=&\frac{1}{\sqrt{2N_k}}
\left(
\begin{array}{c}
0\\
\gamma_{1+}(\vec k)\\
0\\
\gamma_{1-}(\vec k)\\
\end{array}
\right),
\eea
and
\bea
|A_2(\vec k)\rangle&=&\frac{1}{\sqrt{2N_k}}
\left(
\begin{array}{c}
\gamma_{2-}(\vec k)\\
0\\
\gamma_{2+}(\vec k)\\
0
\end{array}
\right), \nn \\
|B_2(\vec k)\rangle&=&\frac{1}{\sqrt{2N_k}}
\left(
\begin{array}{c}
0\\
\gamma_{2+}^*(\vec k)\\
0\\
\gamma_{2-}^*(\vec k)
\end{array}
\right)
\label{eq:newbase}
\eea
where
\bea
\gamma_{1\pm}(\vec k)&=&\sum_{i=1}^3 e^{i\vec k \cdot \hat e_i \pm2 i\theta_i},  \ \ \,
\gamma_{2\pm}(\vec k)=\sum_{i=1}^3 e^{i\vec k \cdot \hat e_i \pm i\theta_i}, \nn\\
N(\vec k)&=&3-\eta_c(\vec k).
\eea
In terms of this set of new bases, $H_\uparrow(\vec k)$ is represented as
\begin{widetext}
\bea
H_\uparrow(\vec k)=
\left[
\begin{array}{cccc}
m-n(\vec k)& -\frac{3}{2} t_\pp & h(\vec k)& 0 \\
-\frac{3}{2}t_\pp & -m +n(\vec k) &0 & h(-\vec k)\\
h^*(\vec k)& 0 & m +n(\vec k)& -\frac{1}{2} t_\pp l^*(\vec k)\\
0& h^*(-\vec k) & -\frac{1}{2} t_\pp l(\vec k)& -m - n(\vec k)
\end{array}
\right],
\label{eq:newham}
\eea
\end{widetext}
where for simplicity $t_\perp$ is set to 0; $n(\vec k)$, $l(\vec k)$,
and $h(\vec k)$ are expressed as
\bea
n(\vec k)&=&\frac{\sqrt 3 \lambda}{N_k} \eta_s(\vec k), \ \ \, \ \ \,
l(\vec k)= \sum_i e^{i\vec k \cdot \hat e_i}, \nn \\
h(\vec k)&=&\frac{i\lambda}{N_k}
\Big\{(\sum_i e^{i\vec k \cdot \hat e_i})^2
-3(\sum_i e^{-i\vec k \cdot \hat e_i})\Big\}.
\eea

In the absence of SO coupling, $h(\vec k)=n(\vec k)=0$,
the above matrix of $H_\uparrow(\vec k)$ is already block diagonalized.
The left-up block represents the Hamiltonian matrix in the subspace
spanned by the bottom band $|\phi_1(\vec k)\rangle$ and
top band $|\phi_4(\vec k)\rangle$,  and the right-bottom block
represents that in the subspace spanned by the middle two bands
$|\phi_{2,3}(\vec k)\rangle$.
Apparently, the bottom and top bands are flat as
\bea
E_{1,4}=\pm\sqrt{(\frac{3}{2}t_\pp)^2+m^2},
\label{eq:flat}
\eea
whose eigen wave functions are solved as
\bea
\left[\begin{array}{c}
|\phi_1(\vec k)\rangle\\
|\phi_4(\vec k)\rangle
\end{array}
\right]
=\left[ \begin{array}{cc}
\sin\frac{\alpha}{2} & \cos\frac{\alpha}{2} \\
\cos\frac{\alpha}{2} & -\sin\frac{\alpha}{2}
\end{array}
\right]
\left[
\begin{array}{c}
|A_1(\vec k)\rangle\\
|B_1 (\vec k) \rangle
\end{array}
\right],
\eea
where $\alpha=\arctan \frac{3t_\pp}{2m}$.
As for the middle two bands, the
spectra can be easily diagonalized as
\bea
E_{2,3}(\vec k)=\pm
\sqrt{\frac{1}{4}t^2_\pp\eta^2_c(\vec k)+m^2}.
\eea
The spectrum is the same as that in graphene at $m=0$.
The eigen wave functions are enriched by orbital structures
which can be solved as
\bea
\left[\begin{array}{c}
|\phi_2(\vec k)\rangle\\
|\phi_3(\vec k)\rangle
\end{array}
\right]
=\left[ \begin{array}{cc}
\sin\frac{\beta}{2} & \cos\frac{\beta}{2} e^{i\phi} \\
\cos\frac{\beta}{2} e^{-i\phi} & -\sin\frac{\beta}{2}
\end{array}
\right]
\left[
\begin{array}{c}
|A_2(\vec k)\rangle\\
|B_2 (\vec k) \rangle
\end{array}
\right], \nn \\
\label{eq:middle}
\eea
where $\beta(\vec k)=\arctan [\frac{t_\pp}{2m} l(\vec k)]$
and $\phi(\vec k)=\arg{l(\vec k)}$.

\subsection{Appearance of flat bands}
According to the analytical solution of spectra Eq.~(\ref{eq:spectra}),
flat bands appear in two different situations:
(i) In the absence of SO coupling such that the bottom
and top bands are flat with the eigen energies described by Eq.~(\ref{eq:flat});
(ii) in the presence of SO coupling,
at $\lambda=\frac{3}{4}t_\pp$, the two middle bands are flat with
the energies $E_{2,3}(\vec k)=\pm \frac{3}{4}t_\pp$.
In both cases, the band flatness implies that we can construct
eigenstates localized in a single hexagon plaquette.
The localized eigenstates for the case of $\lambda=0$ are
constructed in Ref.~\onlinecite{wu2007}, and those for the
case of $\lambda=\frac{3}{4}t_\pp$ were presented
in Ref.~\onlinecite{zhang2011a}.
Since the kinetic energy is suppressed in the flat bands,
interaction effects are nonperturbative.
Wigner crystallization \cite{wu2007} and ferromagnetism
\cite{zhang2010} have been studied in the flat band at
$\lambda=0$.

\section{Band topology and band crossings}
\label{sect:topo}

\begin{figure}
\centering\includegraphics[width=0.9\linewidth]{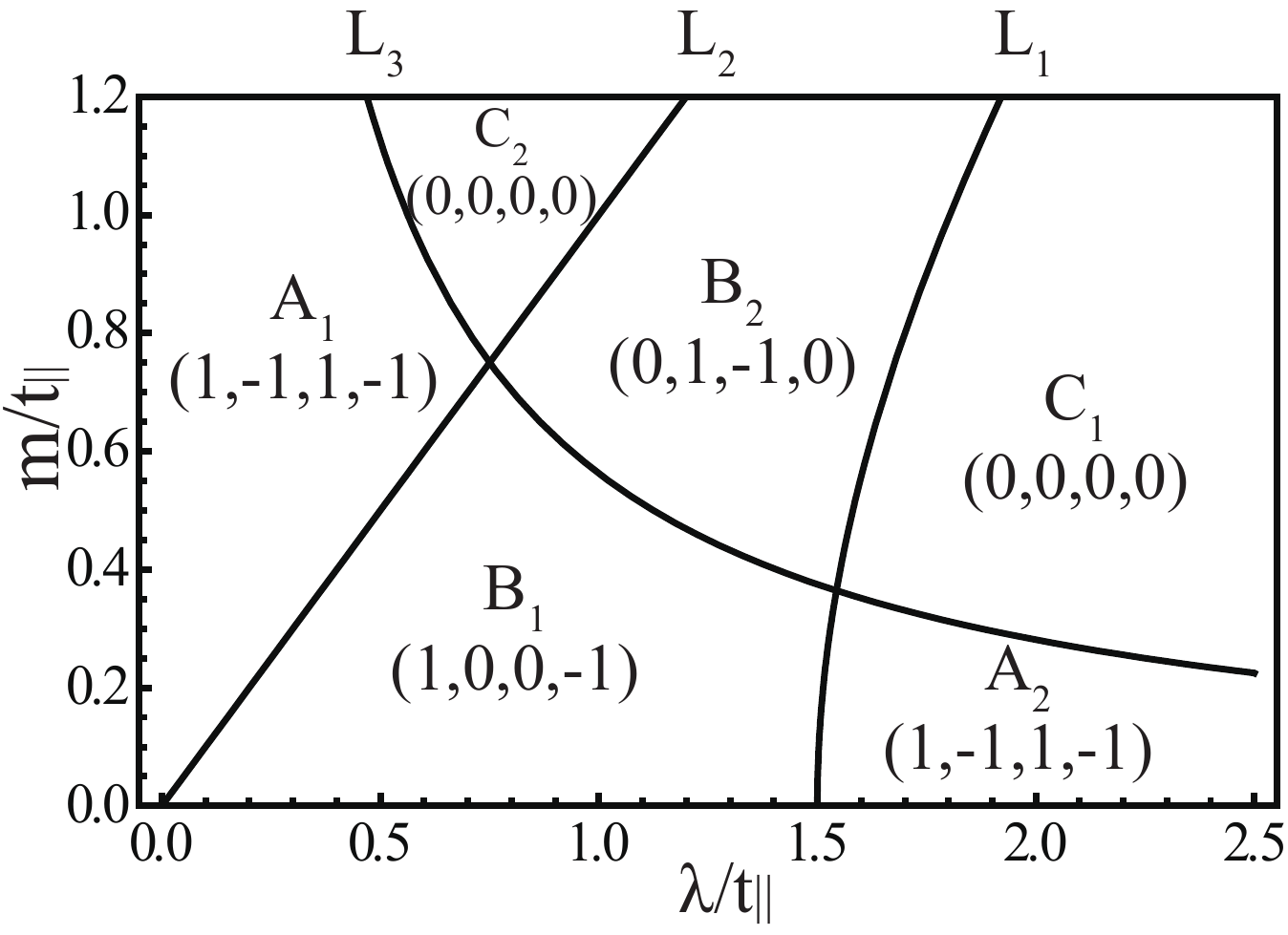}
\caption{Phases with different spin Chern number patterns
$(C_{s1}, C_{s2}, C_{s3}, C_{s4})$ vs
SO coupling strength $\lambda$ and the sub-lattice asymmetry parameter $m$.
Due to the $s_z$ conservation and TR symmetry, only those of the
four $s=\uparrow$ bands are shown.
Phase boundaries $L_{1,2,3}$ satisfy the level crossing conditions
located at $\Gamma$, $K$, and $K^\prime$, respectively.
Their analytic expressions are $\lambda^2-m^2=(\frac{3}{4}t_\pp)^2$,
$\lambda=m$, and $\lambda\,m=(\frac{3}{4}t_\pp)^2$, respectively.
$L_1$ and $L_3$ intersect at $(\lambda,m)=(\frac{3}{4}(\sqrt{5}+2)
,\frac{3}{4}(\sqrt{5}-2))\approx (1.54,0.36)$,
and $L_2$ and $L_3$ intersect at $(\lambda, m)=(\frac{3}{4},
\frac{3}{4})$.
}
\label{fig:phase}
\end{figure}

In this section, we study the topology of band structures after SO
coupling $\lambda$ is turned on.
Due to the $s_z$ conservation, the $Z_2$ topological class is augmented
to the spin Chern class.
Without loss of generality, we only use the pattern of Chern numbers of
the sector $s=\uparrow$ to characterize the band topology, and that
of the $s=\downarrow$ sector is just with an opposite sign.
The Berry curvature for the $i$-th band is defined as
\bea
F_i(\vec k)=\partial_{k_x} A_y(\vec k) -\partial_{k_y} A_x(\vec k)
\eea
in which the Berry connection is defined as $\vec A_i(\vec k)=
-i \avg{\phi_i(\vec k)|\vec \nabla_k|\phi_i(\vec k)}$.
The spin Chern number of band $i$ can be obtained through the
integral over the entire first Brillouin zone as
\bea
C_{s,i}=\frac{1}{2\pi}\int_{FBZ} d k_x dk_y F_i(\vec k_x, \vec k_y).
\label{eq:spinchern}
\eea

\begin{figure*}
\centering\includegraphics[width=0.25\textwidth]{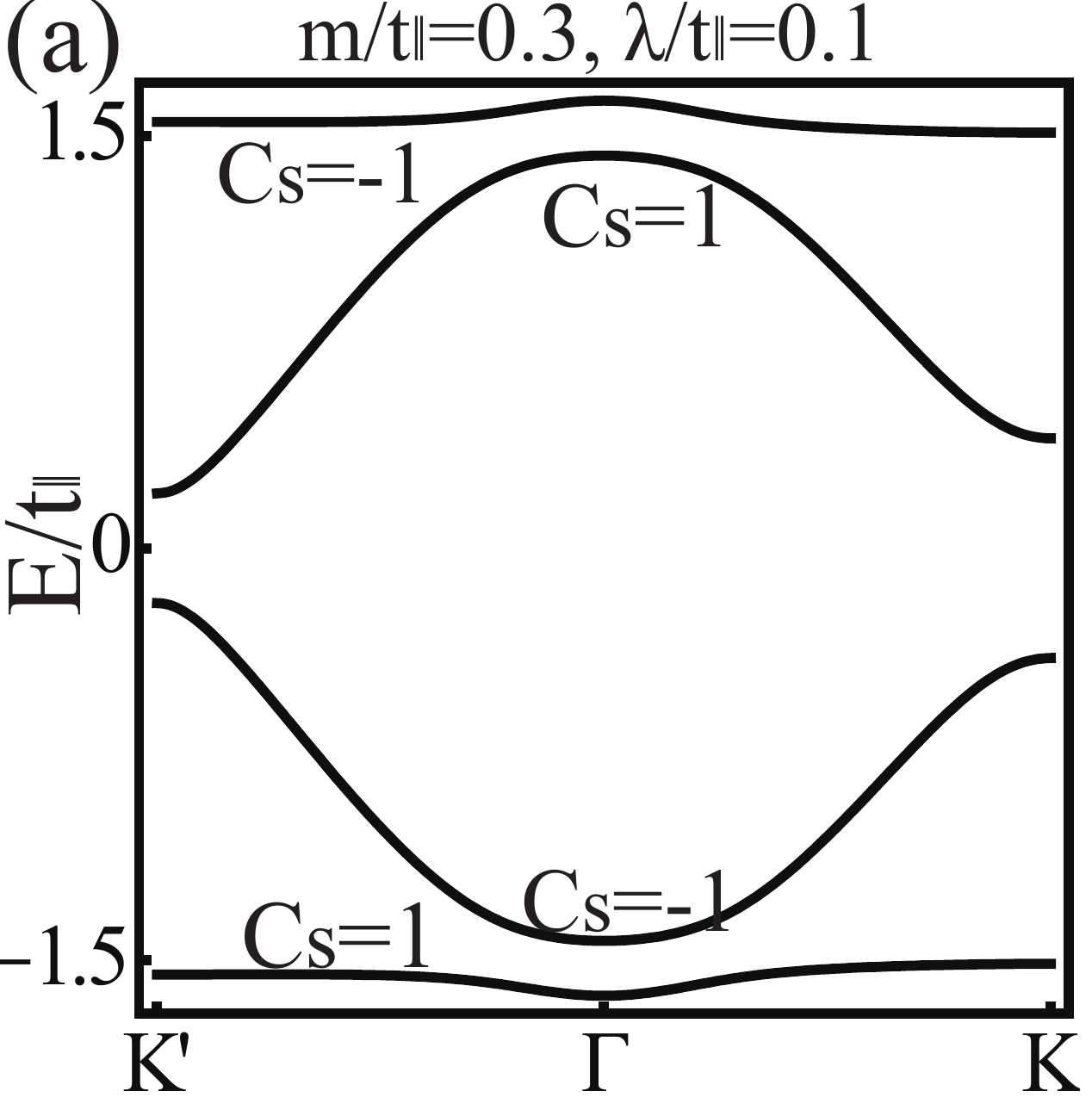}
\centering\includegraphics[width=0.25\textwidth]{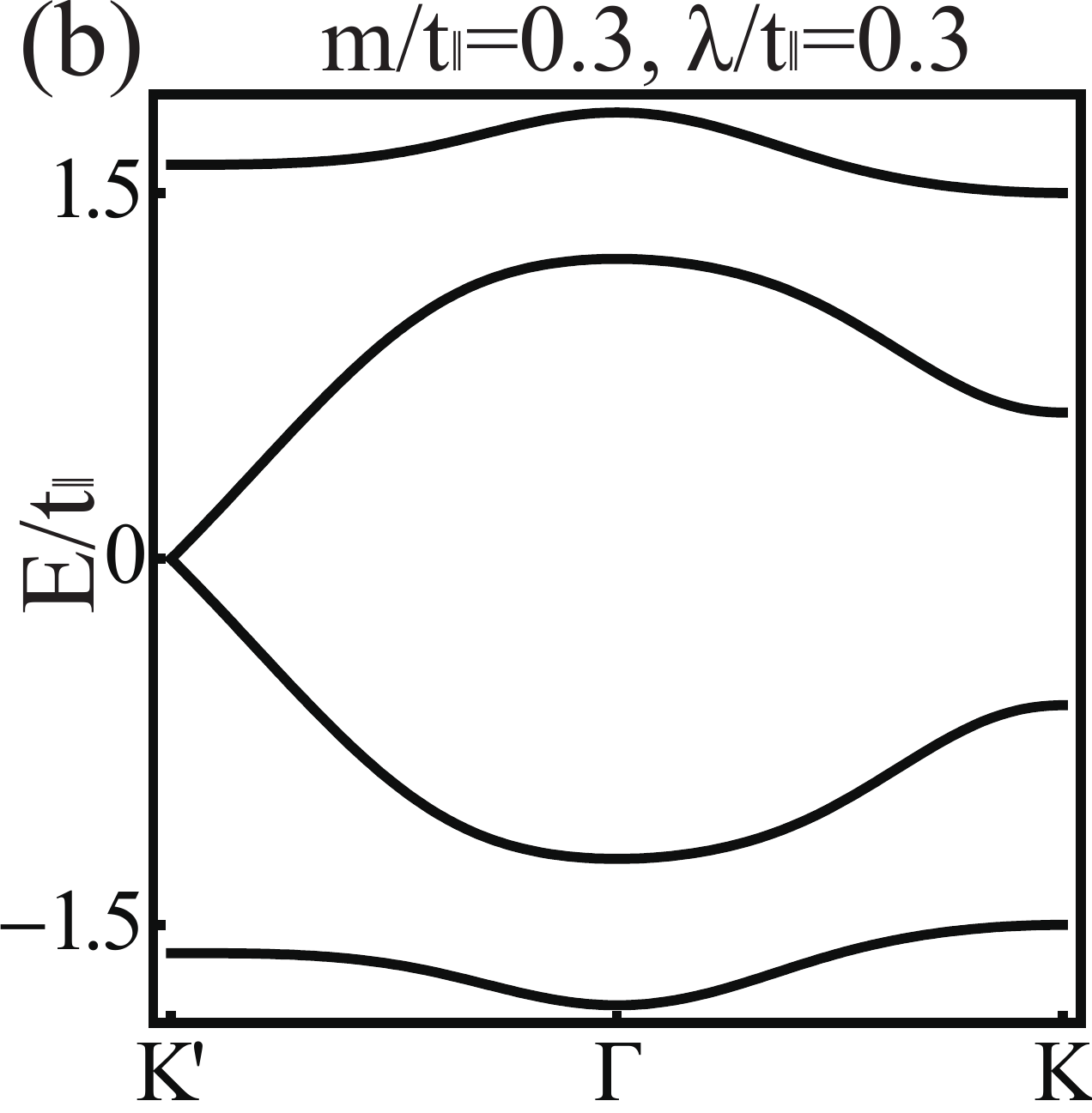}
\centering\includegraphics[width=0.25\textwidth]{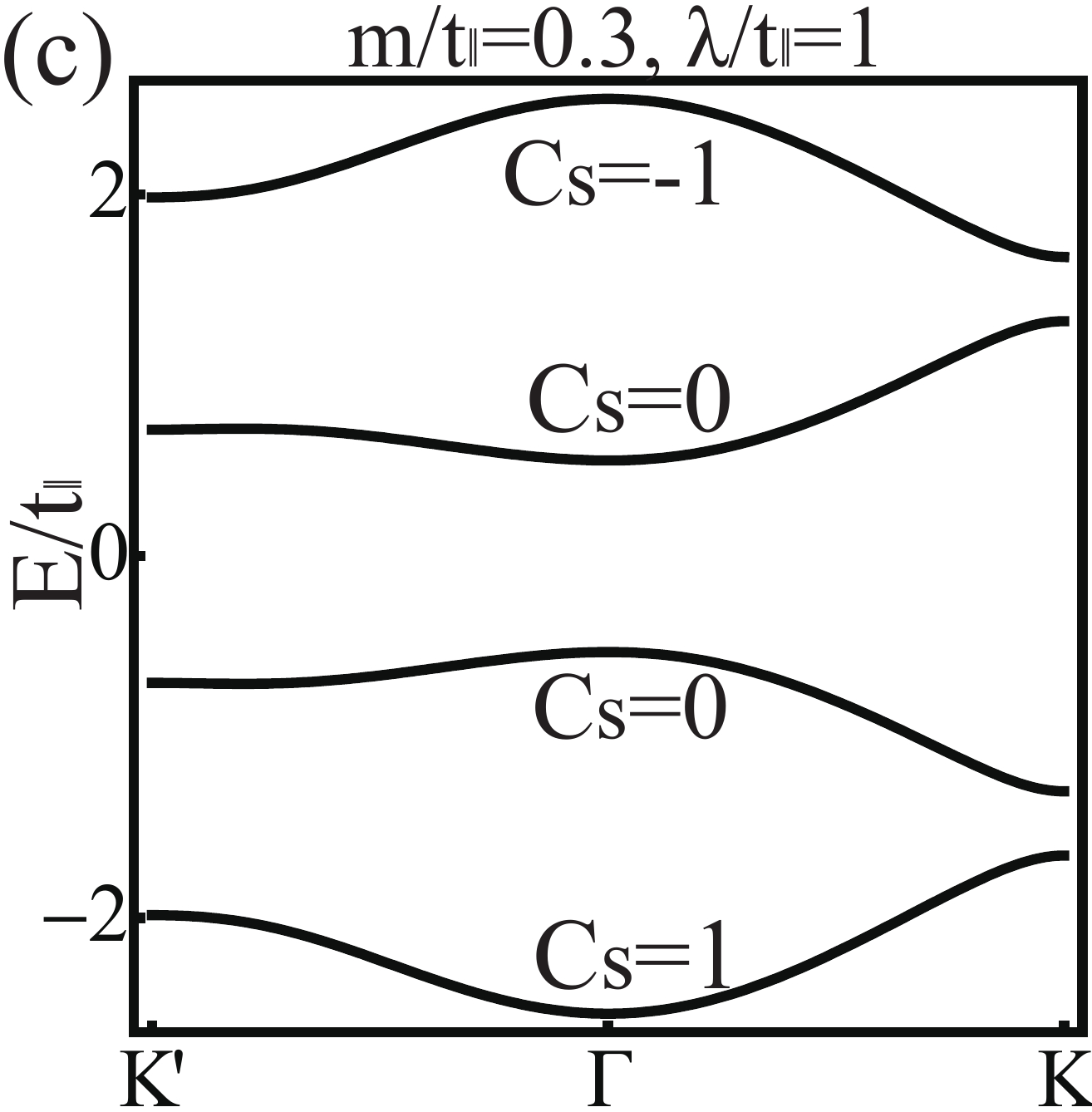}
\centering\includegraphics[width=0.25\textwidth]{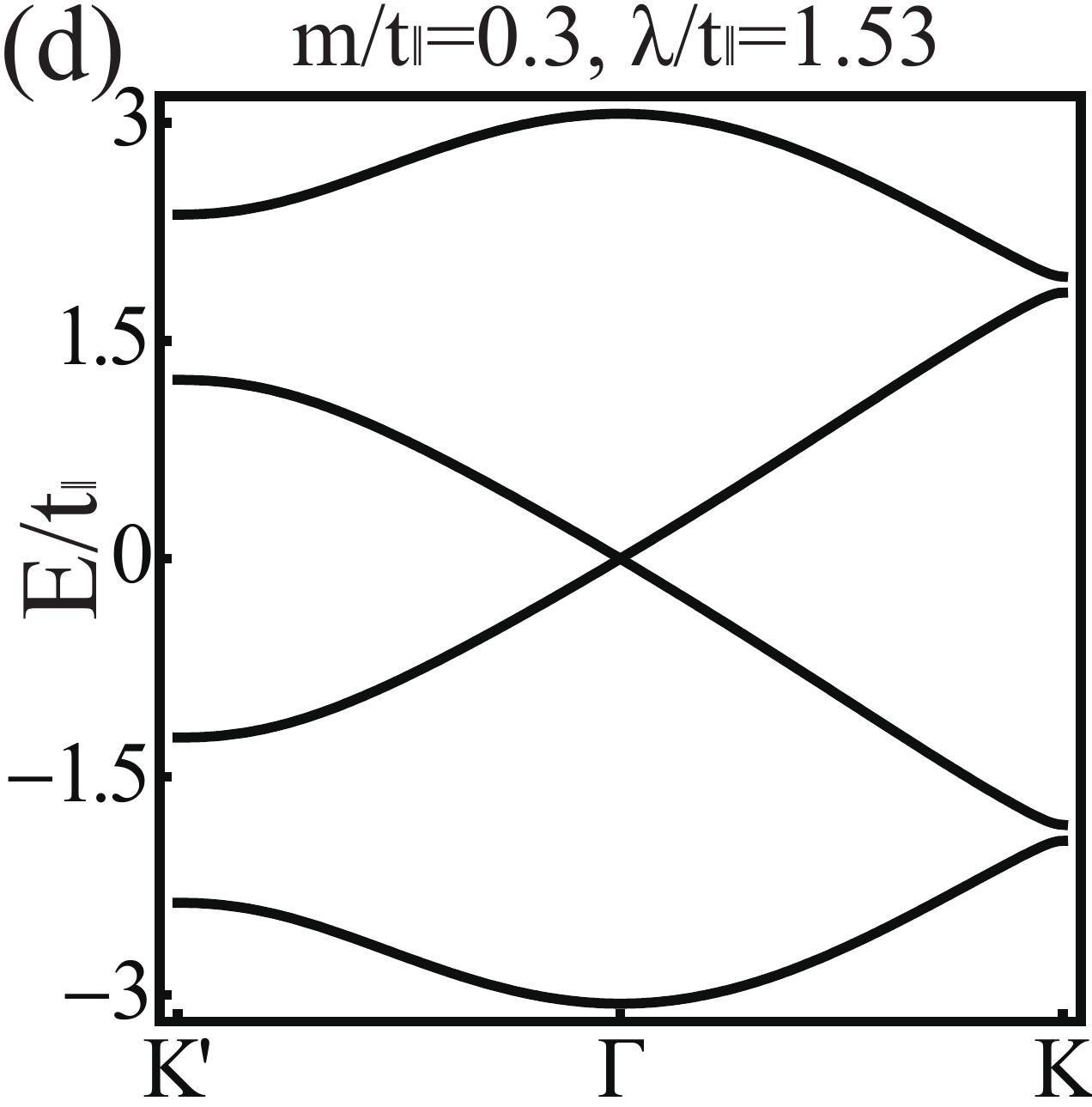}
\centering\includegraphics[width=0.25\textwidth]{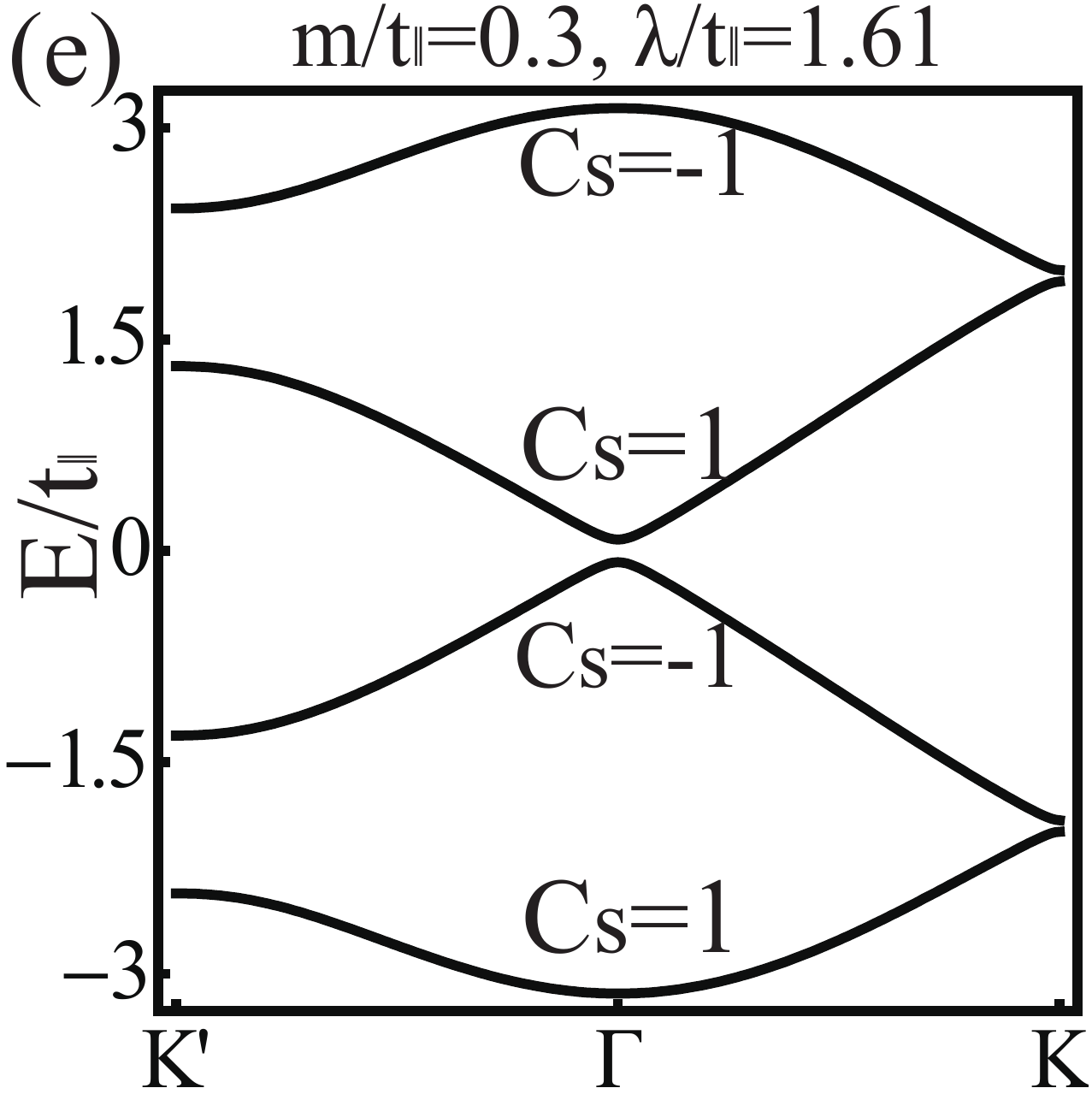}
\centering\includegraphics[width=0.25\textwidth]{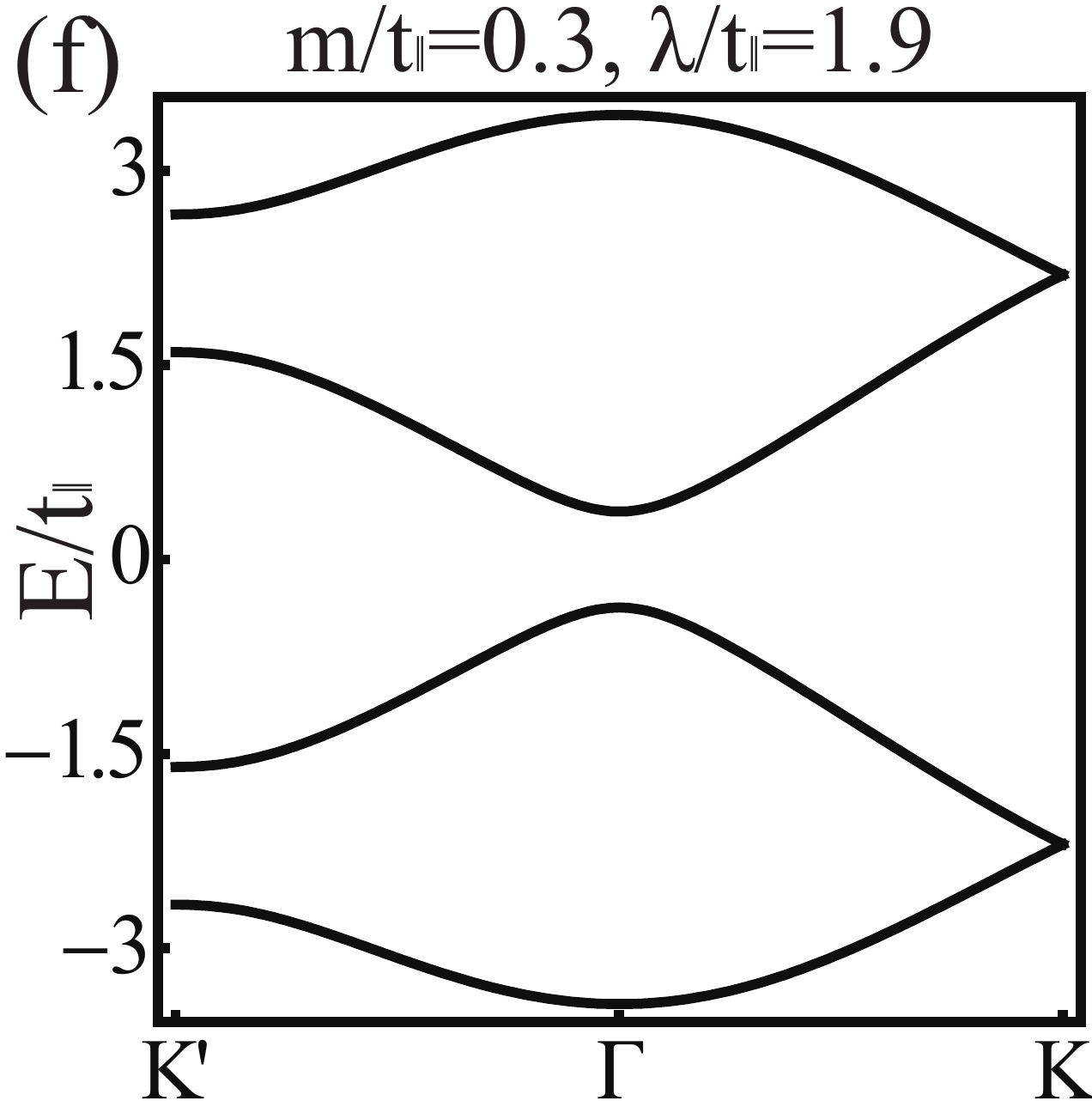}
\centering\includegraphics[width=0.25\textwidth]{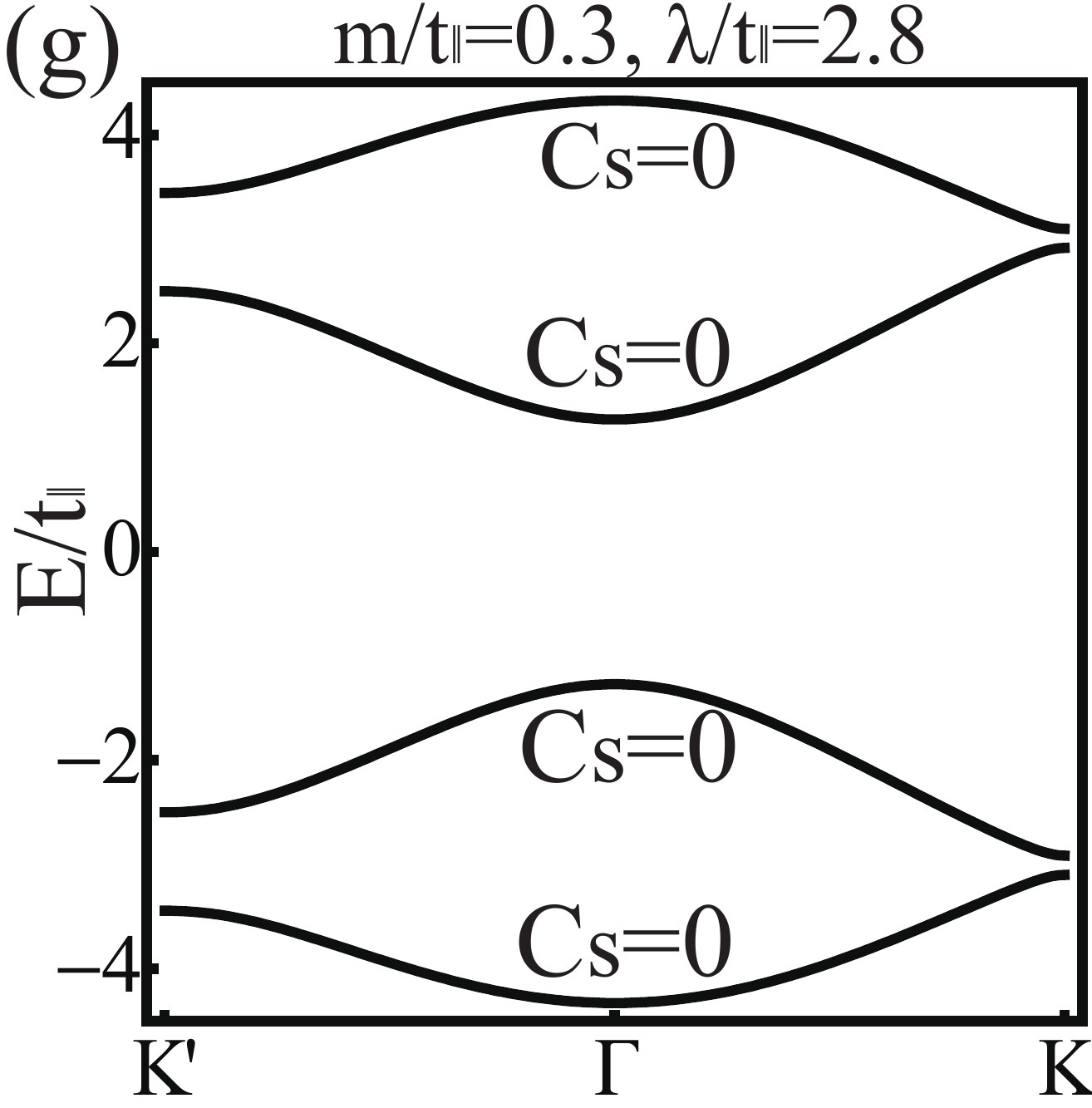}
\caption{The spectra along the cut of $K^\prime$-$\Gamma$-$K$ in Brillouin
zone.
The spectra evolution is shown with fixed $m/t_{\pp}=0.3$ and increasing $\lambda$
from 0.1 (a) to 2.8 (g), which passes phases $A_1$, $B_1$, $A_2$,
and $C_1$.
The pattern of spin-Chern numbers in the gapped states are marked.
Parameters of (b), (d), and (f) are located at phase boundaries
and gaps are closed at $K^\prime$, $\Gamma$, and $K$ points for
(b), (d), and (f), respectively.
Please note the appearance of single Dirac cones for
the sector of spin-$\uparrow$, which is possible
in 2D when two masses from sublattice asymmetry
and SO coupling compete.
}\label{fig:path1}
\end{figure*}

\begin{figure*}
\centering\includegraphics[width=0.25\linewidth]{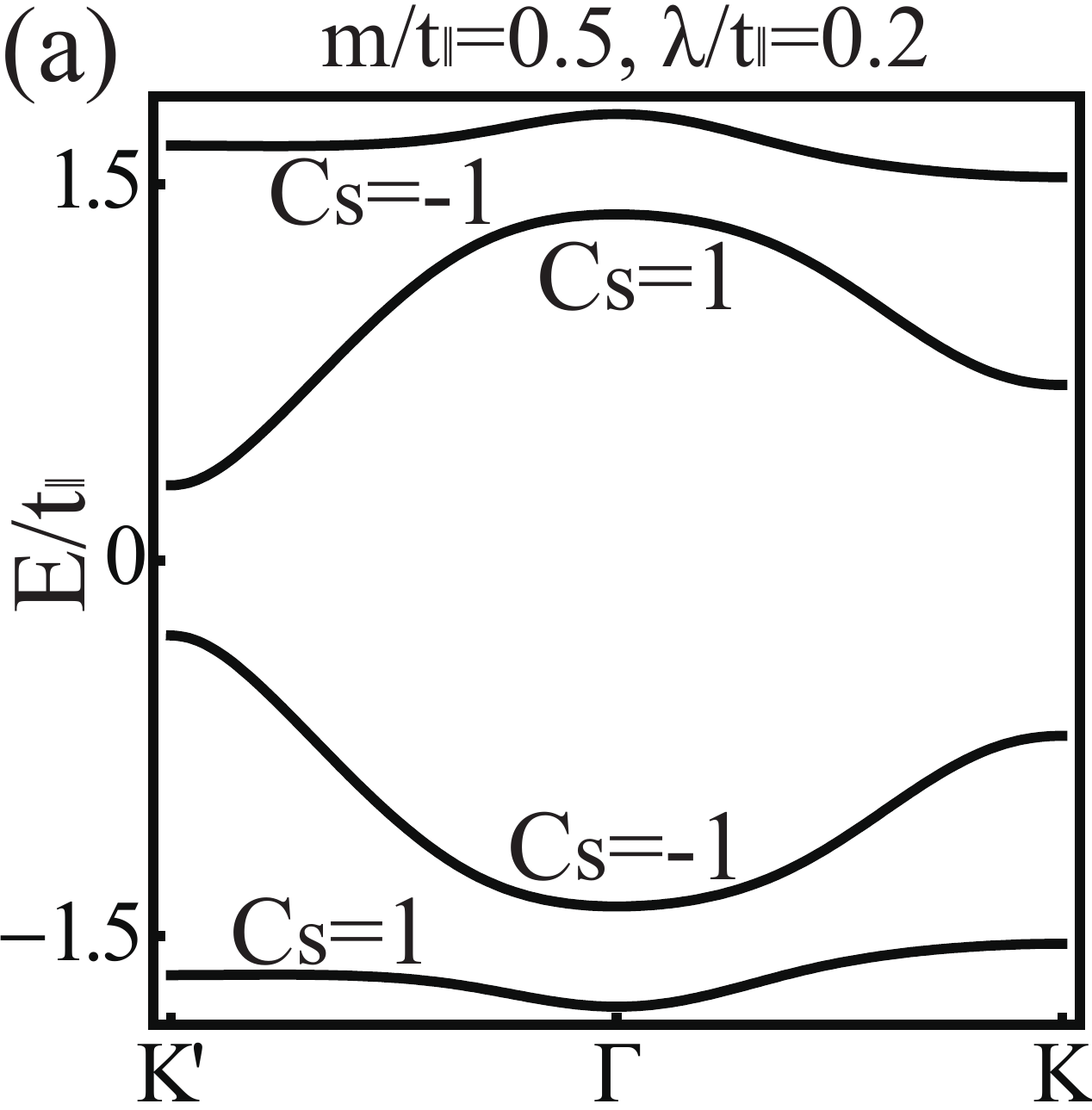}
\centering\includegraphics[width=0.25\linewidth]{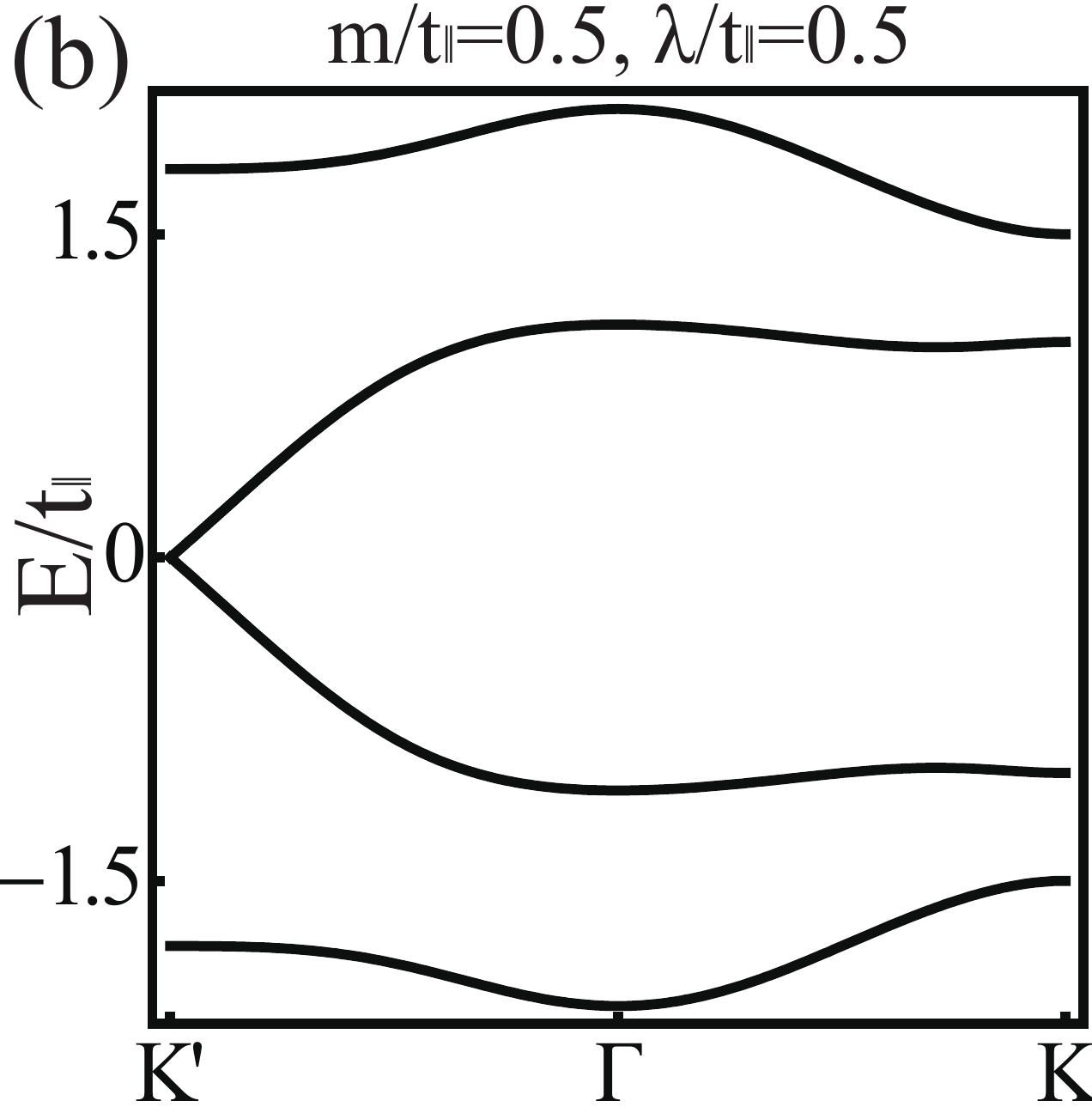}
\centering\includegraphics[width=0.25\linewidth]{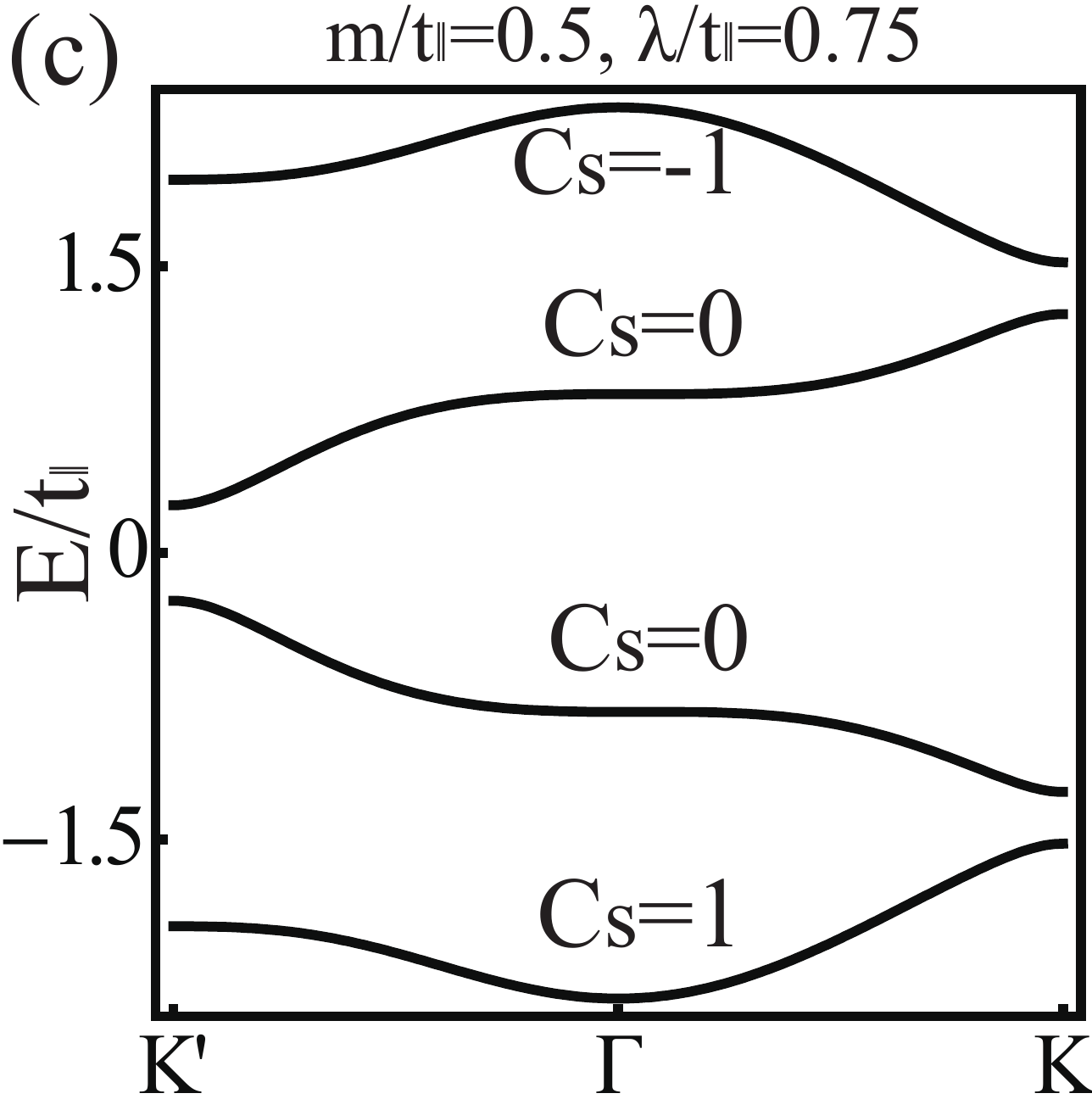}
\centering\includegraphics[width=0.25\linewidth]{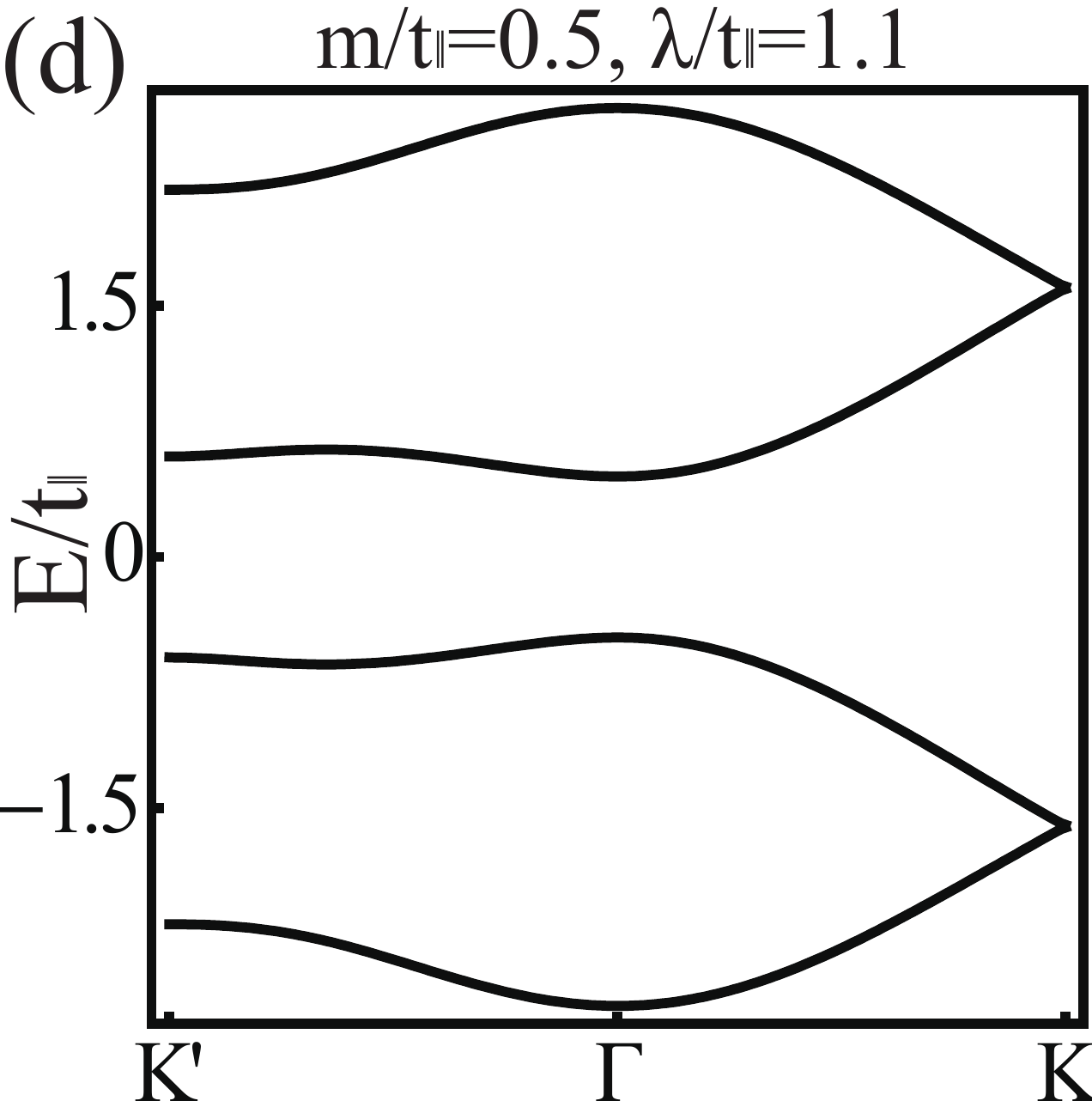}
\centering\includegraphics[width=0.25\linewidth]{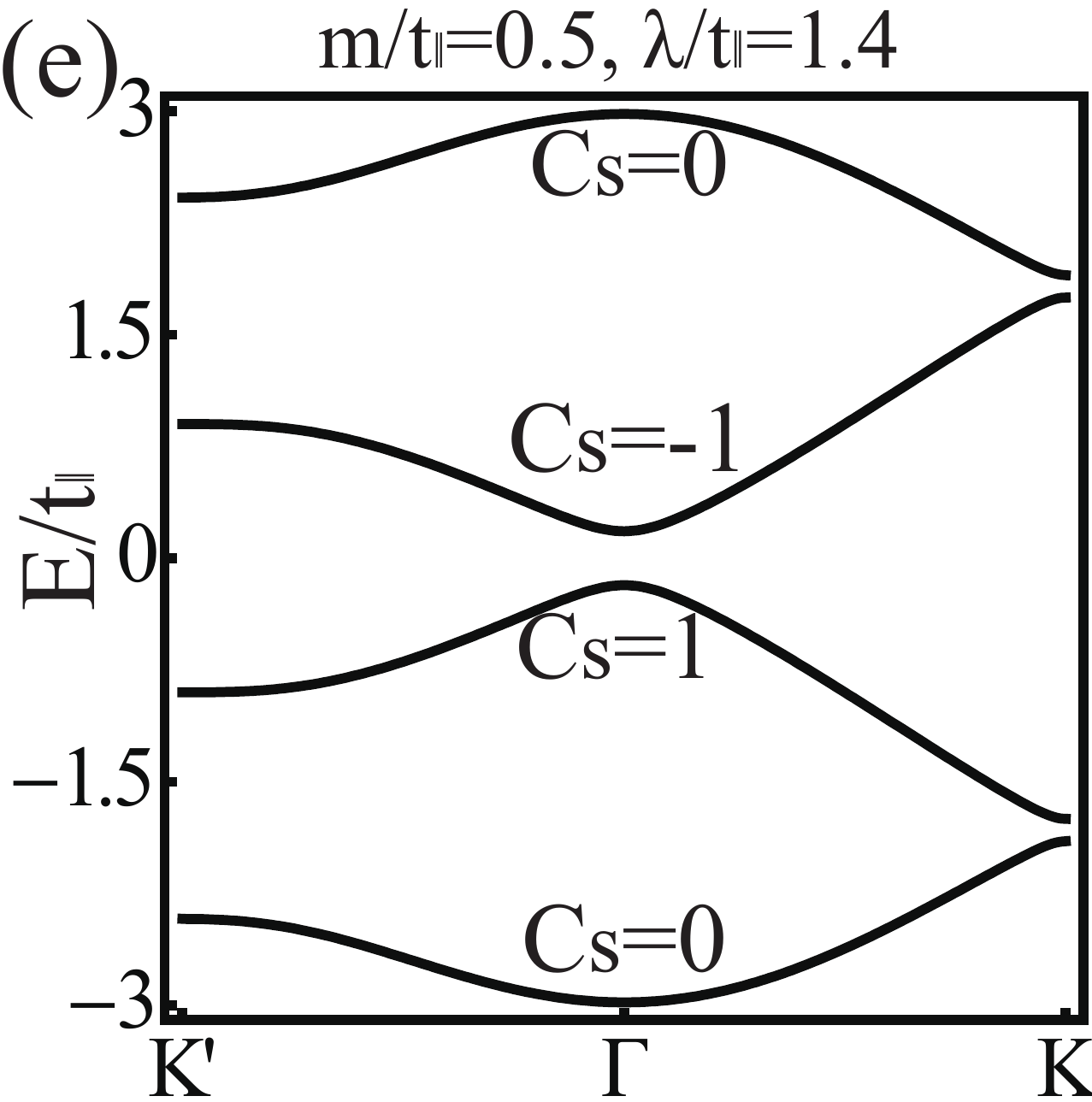}
\centering\includegraphics[width=0.25\linewidth]{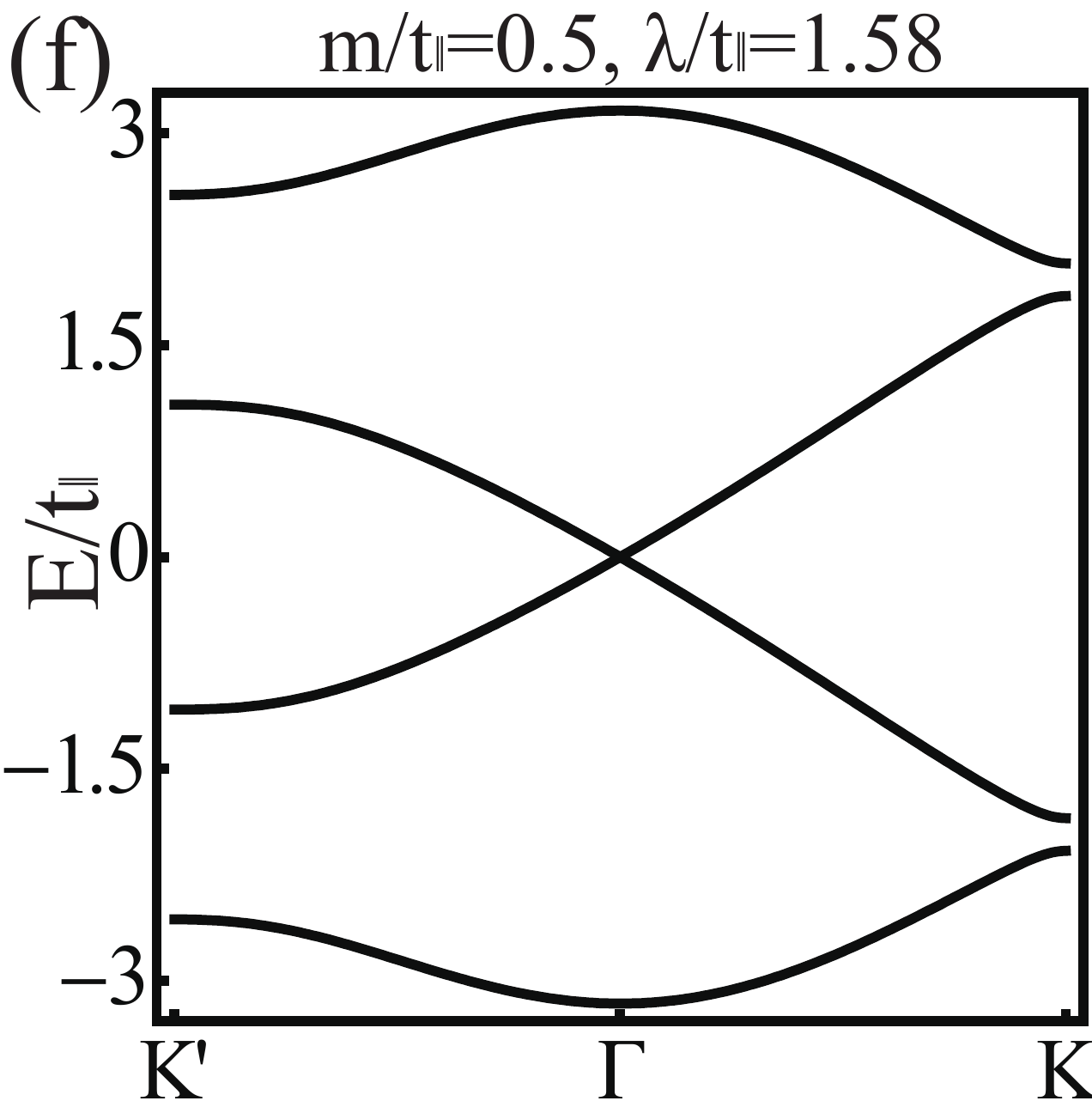}
\centering\includegraphics[width=0.25\linewidth]{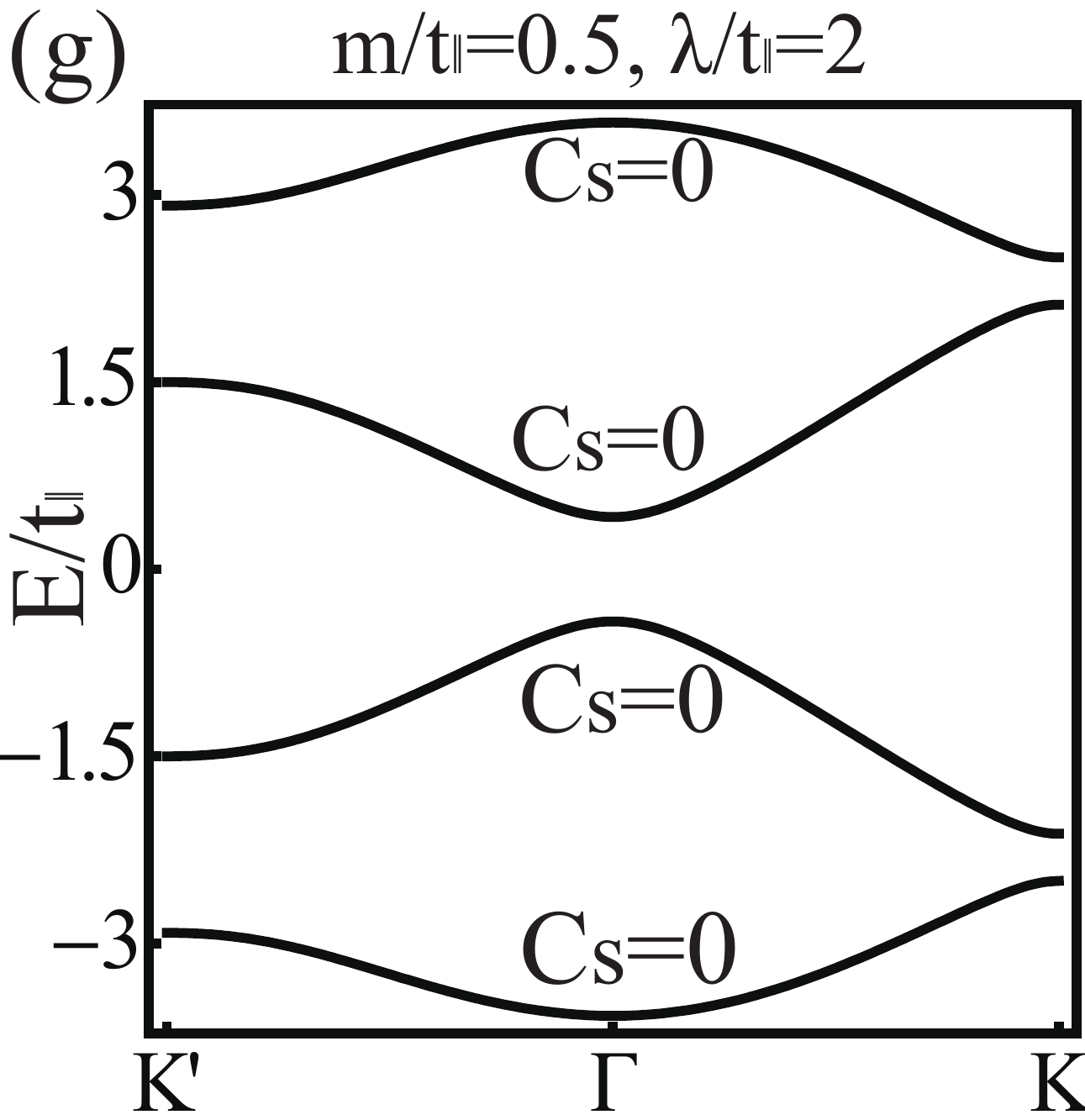}
\caption{The same plot as in Fig. \ref{fig:path1} but for the
evolution with fixed $m/t_{\pp}=0.5$ and increasing $\lambda$
from 0.2 (a) to 2 (g), which passes phases $A_1$, $B_1$, and $B_2$
and $C_1$.
Gaps are closed at $K^\prime$, $K$, and $\Gamma$ points in
(b), (d), and (f), respectively.
}\label{fig:path2}
\end{figure*}

\subsection{Band crossings at $\Gamma$, $K$ and $K^\prime$}
We have performed the numerical integration for spin Chern
numbers $(C_{s,1},C_{s,2},C_{s,3}, C_{s,4})$ for $H_\uparrow(\vec k)$
as presented in Fig.~\ref{fig:phase} based on Eq.~(\ref{eq:spinchern}).
The phase boundary lines $L_{1,2,3}$ are associated with band
touching, which occurs at high symmetry points $\Gamma$, $K$,
and $K^\prime$, respectively.
The momenta of these points are defined as $(0,0)$,
$(\pm\frac{4 \pi}{3\sqrt 3}, 0)$.
Since the dispersions of $H_\uparrow(\vec k)$ are symmetric with respect
to zero energy, the band crossing occurs either between bands 2 and
3 at zero energy, or between 1 and 2, 3 and 4 symmetrically with
respect to zero energy.

We first check the crossing  at the $\Gamma$-point.
According to Eq.~(\ref{eq:spectra}), the energies of the two middle levels are
\bea
E_{2,3}(\Gamma)=\pm \left(\lambda-\sqrt{m^2+\big(\frac{3}{2}t_{\pp}\big)^2}
\right).
\label{eq:Ggap}
\eea
The level crossing can only occur at zero energy with the
hyperbolic condition
\bea
\lambda^2=m^2+\Big(\frac{3}{2}t_\pp\Big)^2,
\label{eq:L1}
\eea
which corresponds to line $L_1$ in Fig. \ref{fig:phase}.

The sublattice asymmetry parameter $m$ and SO coupling $\lambda$ are
different mass generation mechanisms.
The former breaks parity and contributes equally at $K$ and $K^\prime$,
while the latter exhibits opposite signs.
Their total effects superpose constructively or destructively at $K$
and $K^\prime$, respectively, as shown in the spectra of the
two lower energy levels at $K$ and $K^\prime$.
At $K^\prime=(-\frac{4\pi}{3\sqrt 3},0)$, they are
\bea
E_{2,3}(K')=\pm (\lambda-m),
\label{eq:Kpgap}
\eea
and those at $K=(\frac{4\pi}{3\sqrt 3},0)$ are
\bea
E_{1,4}(K)&=&\mp\sqrt{(m-\lambda)^2+\Big(\frac{3}{2}t_\pp\Big)^2}, \nn\\
E_{2,3}(K)&=&\mp(m+\lambda).
\label{eq:Kgap}
\eea
Thus the level crossing at $K^\prime$ occurs at zero energy
with the relation
\bea
\lambda=m,
\label{eq:L2}
\eea
which is line $L_2$ in Fig. \ref{fig:phase}.
Similarly, the level crossing at $K$ occurs when $E_2(K)=E_1(K)$ leading to
the condition
\bea
\lambda\, m=\Big(\frac{3}{4}t_\pp\Big)^2,
\label{eq:L3}
\eea
which is line $L_3$ in Fig.~\ref{fig:phase}.

\begin{figure*}
\centering\includegraphics[width=0.25\linewidth]{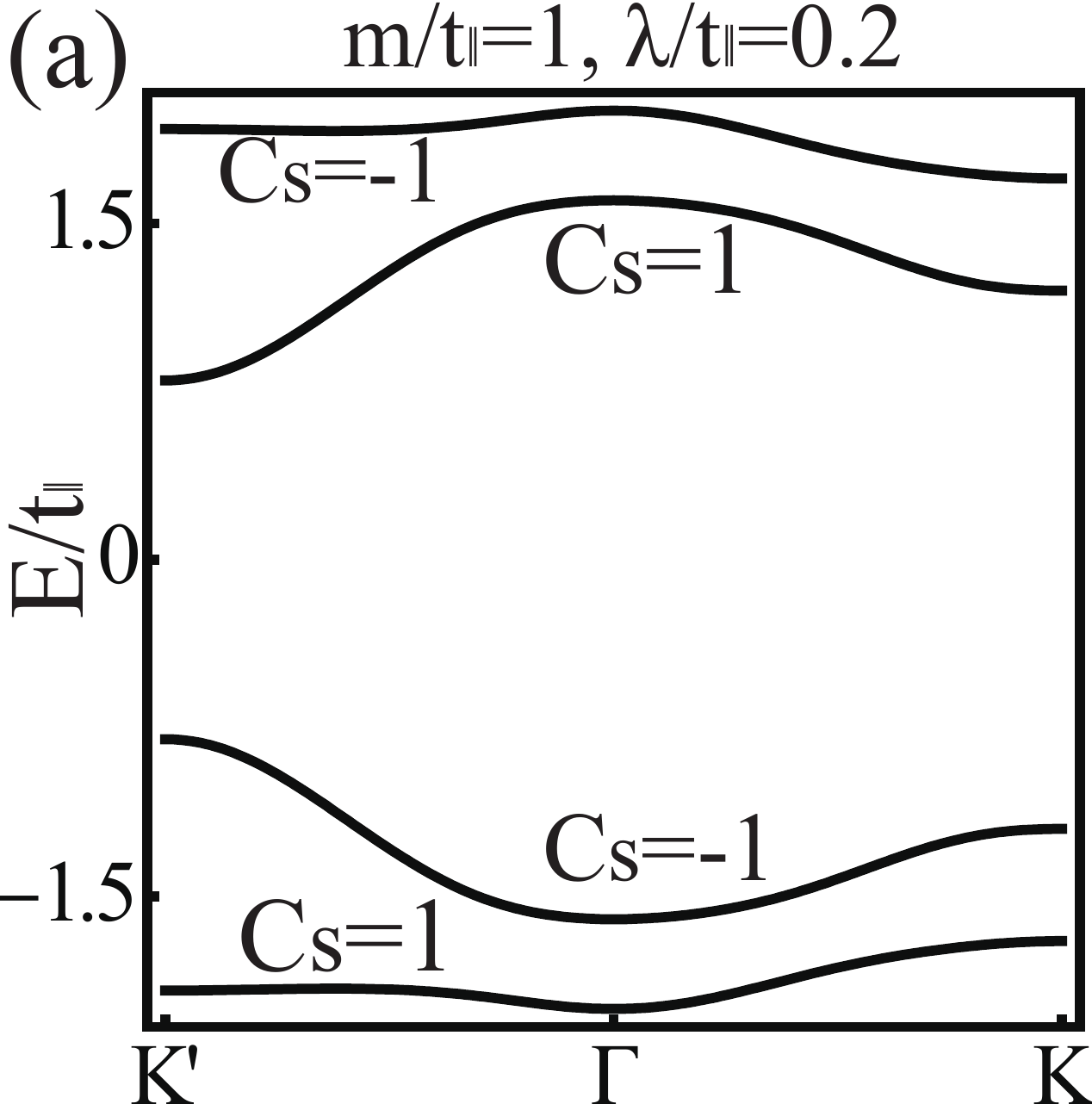}
\centering\includegraphics[width=0.25\linewidth]{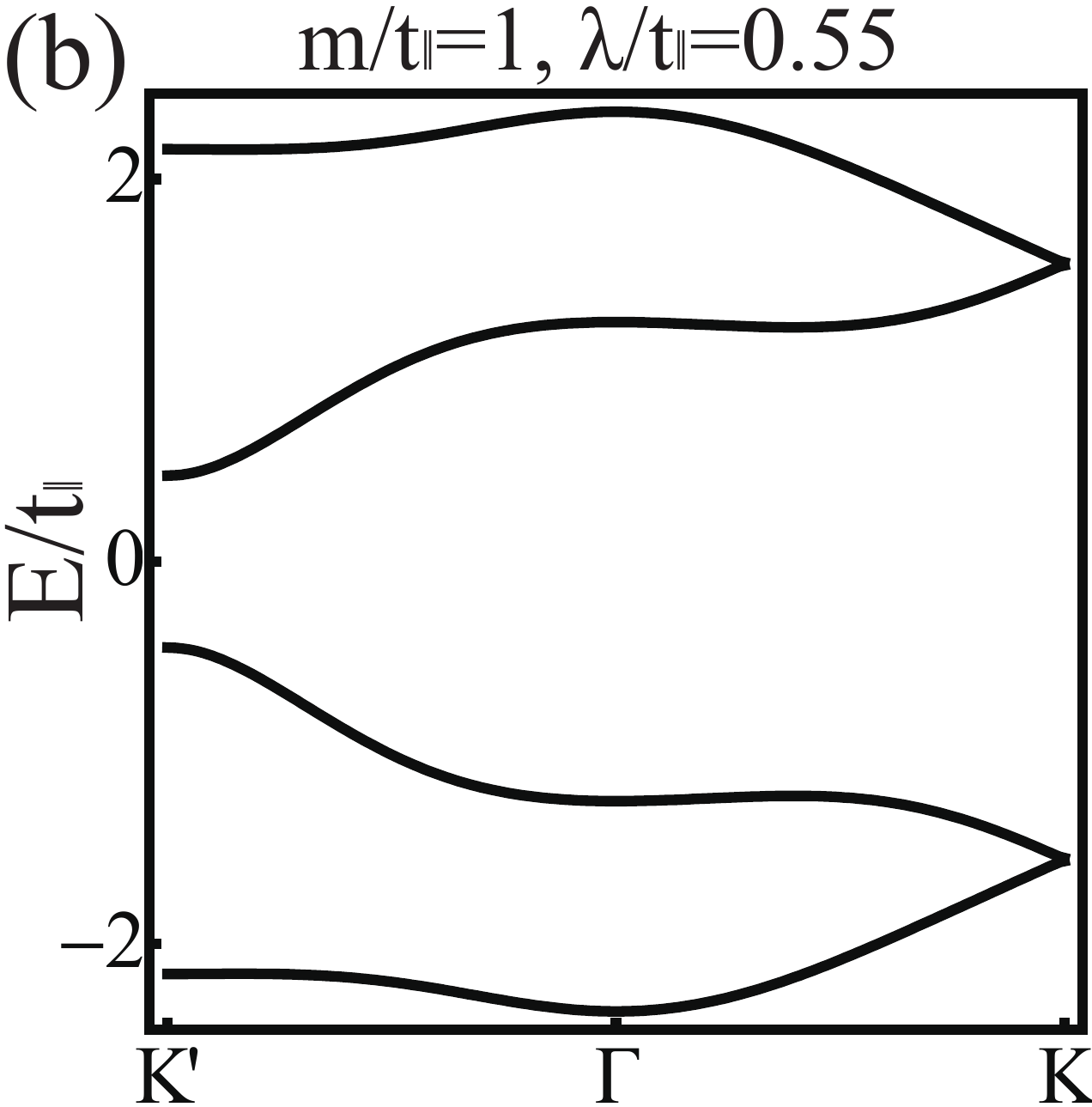}
\centering\includegraphics[width=0.25\linewidth]{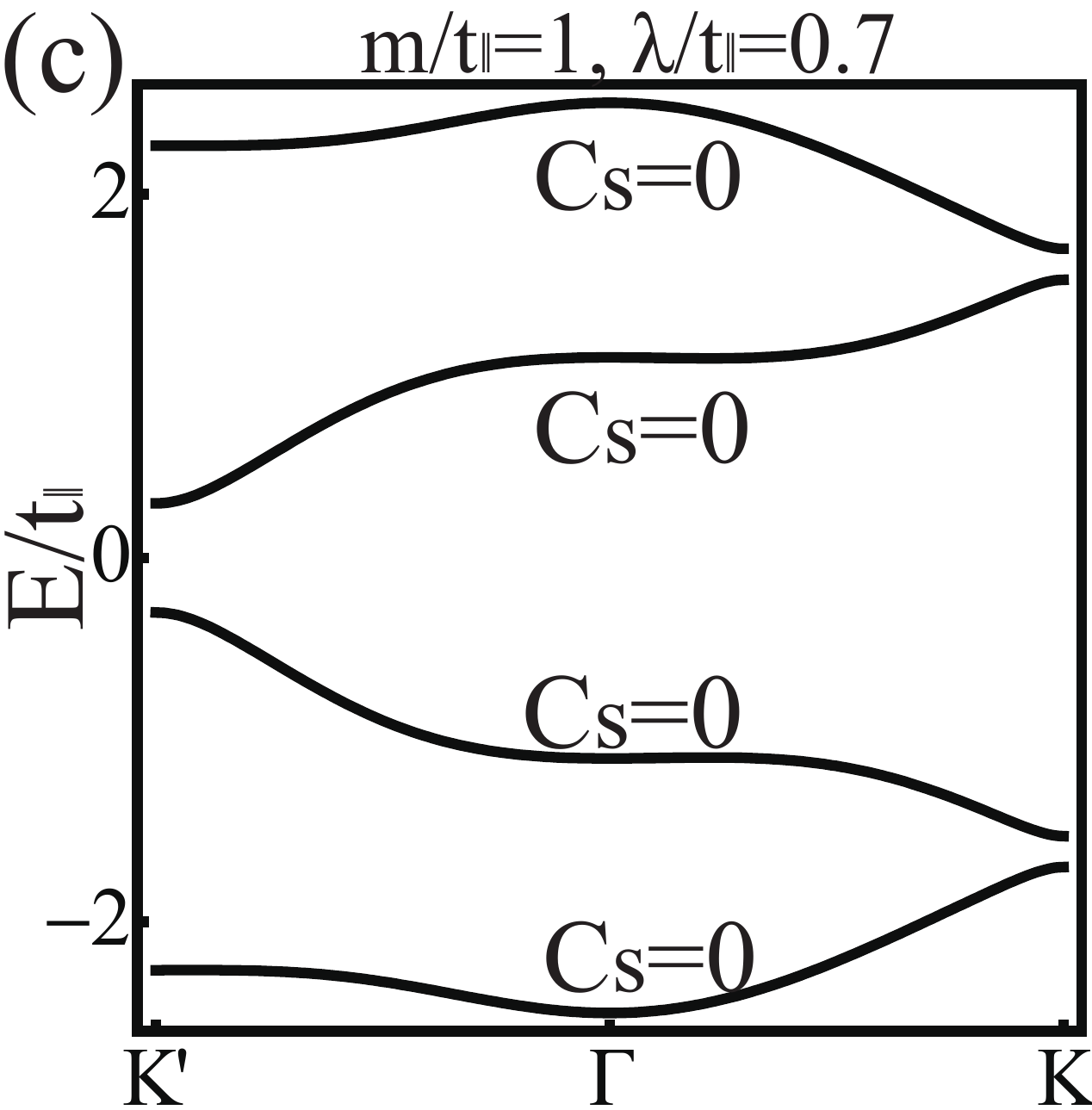}
\centering\includegraphics[width=0.25\linewidth]{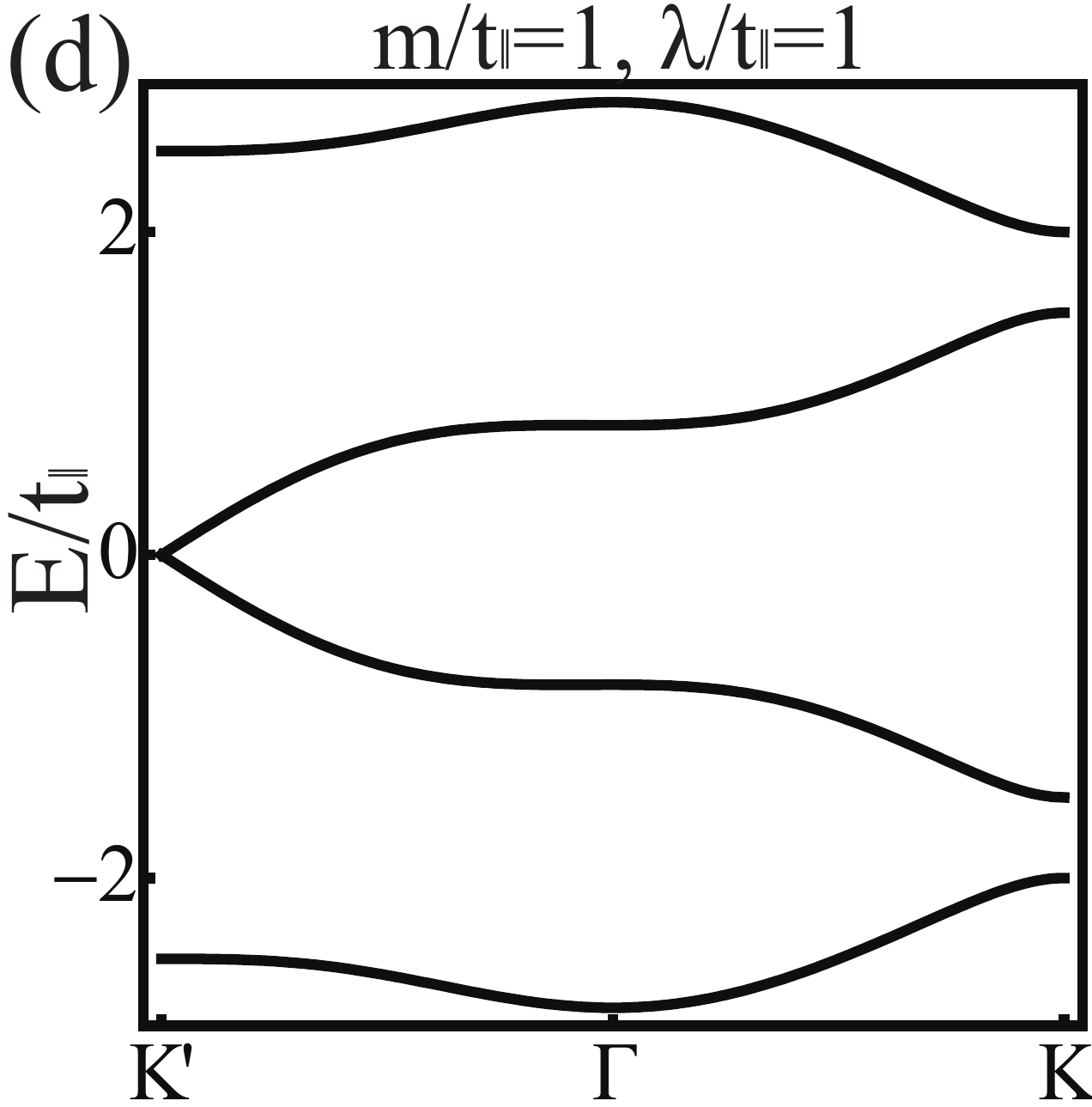}
\centering\includegraphics[width=0.25\linewidth]{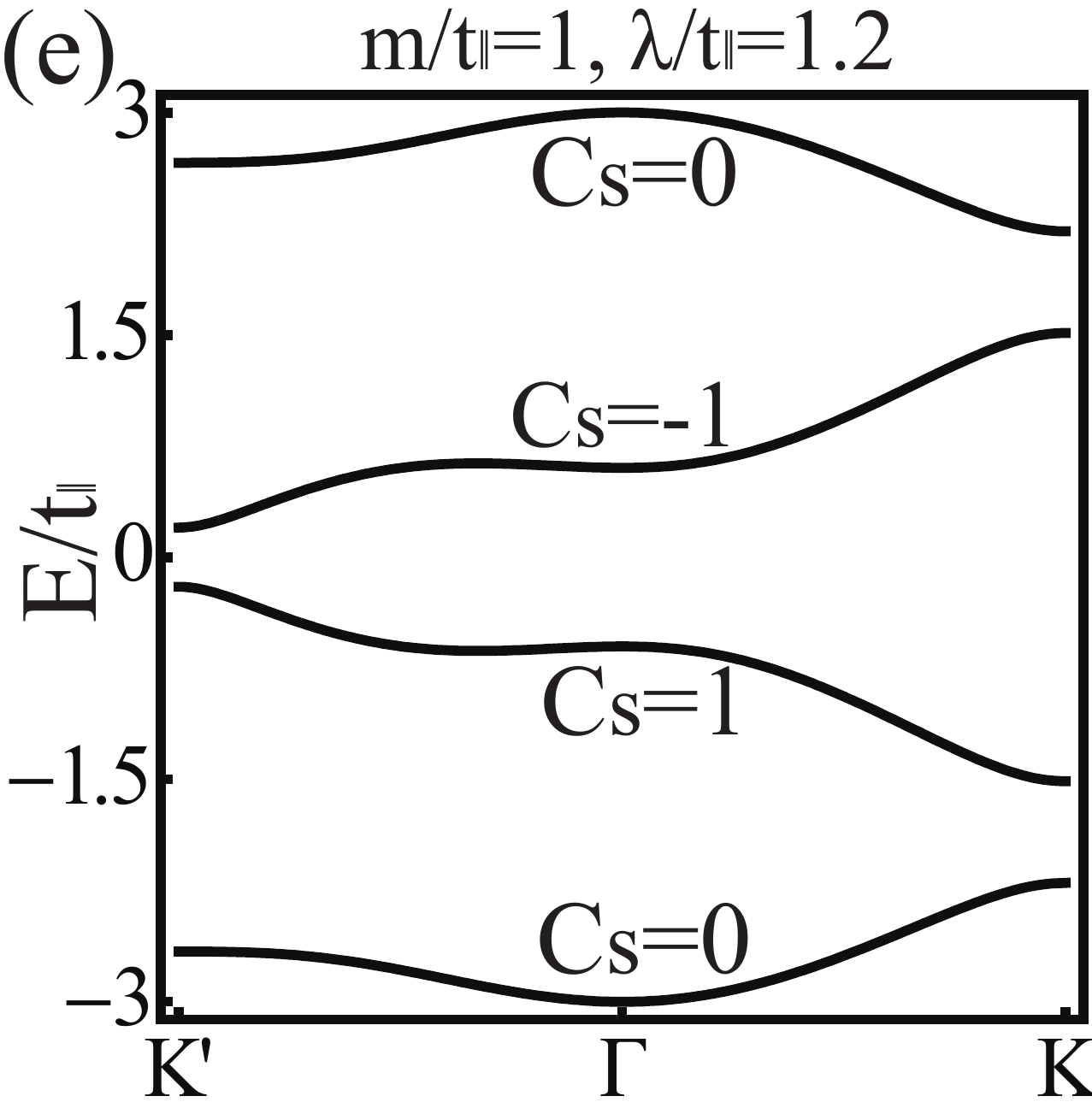}
\centering\includegraphics[width=0.25\linewidth]{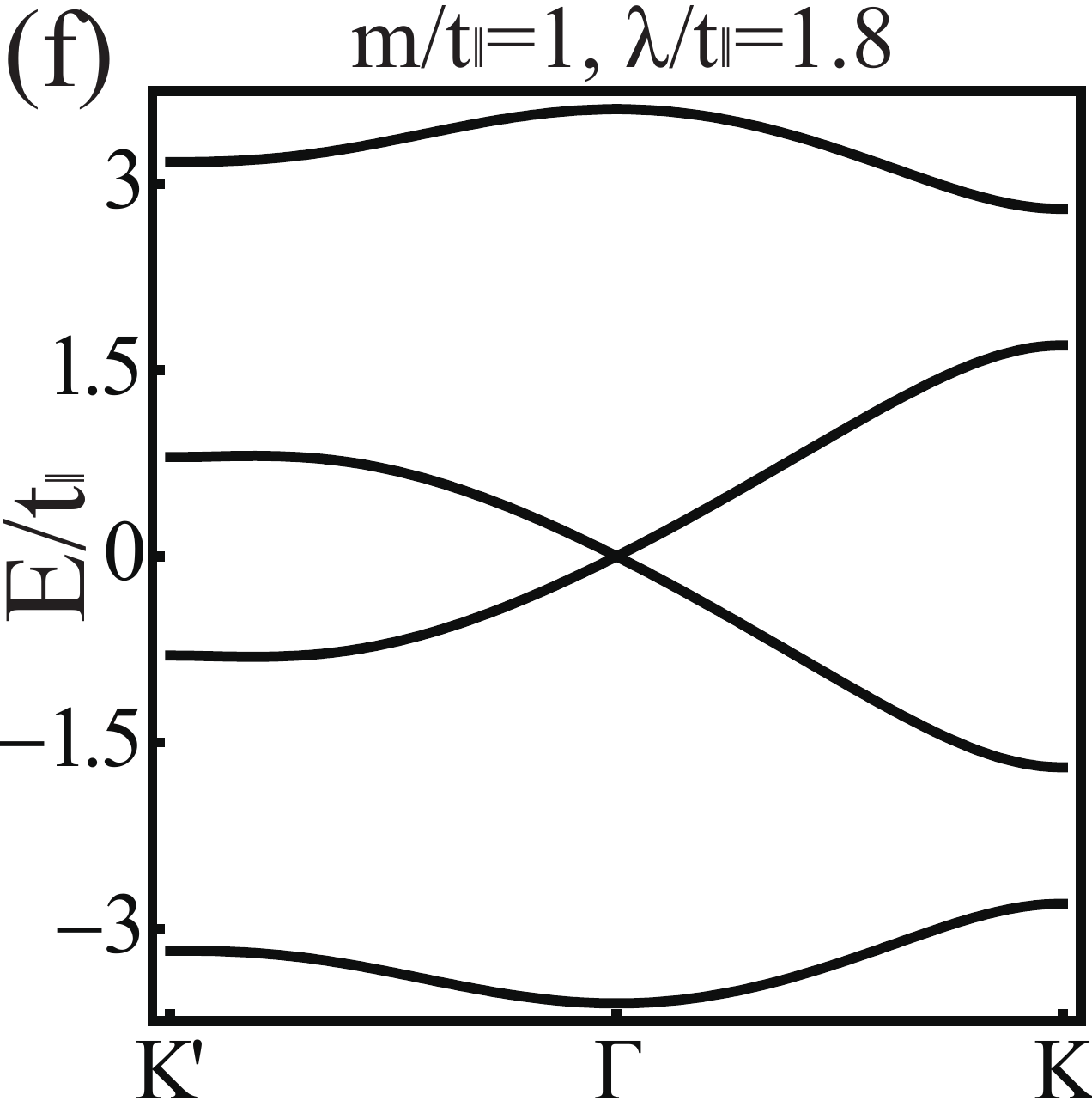}
\centering\includegraphics[width=0.25\linewidth]{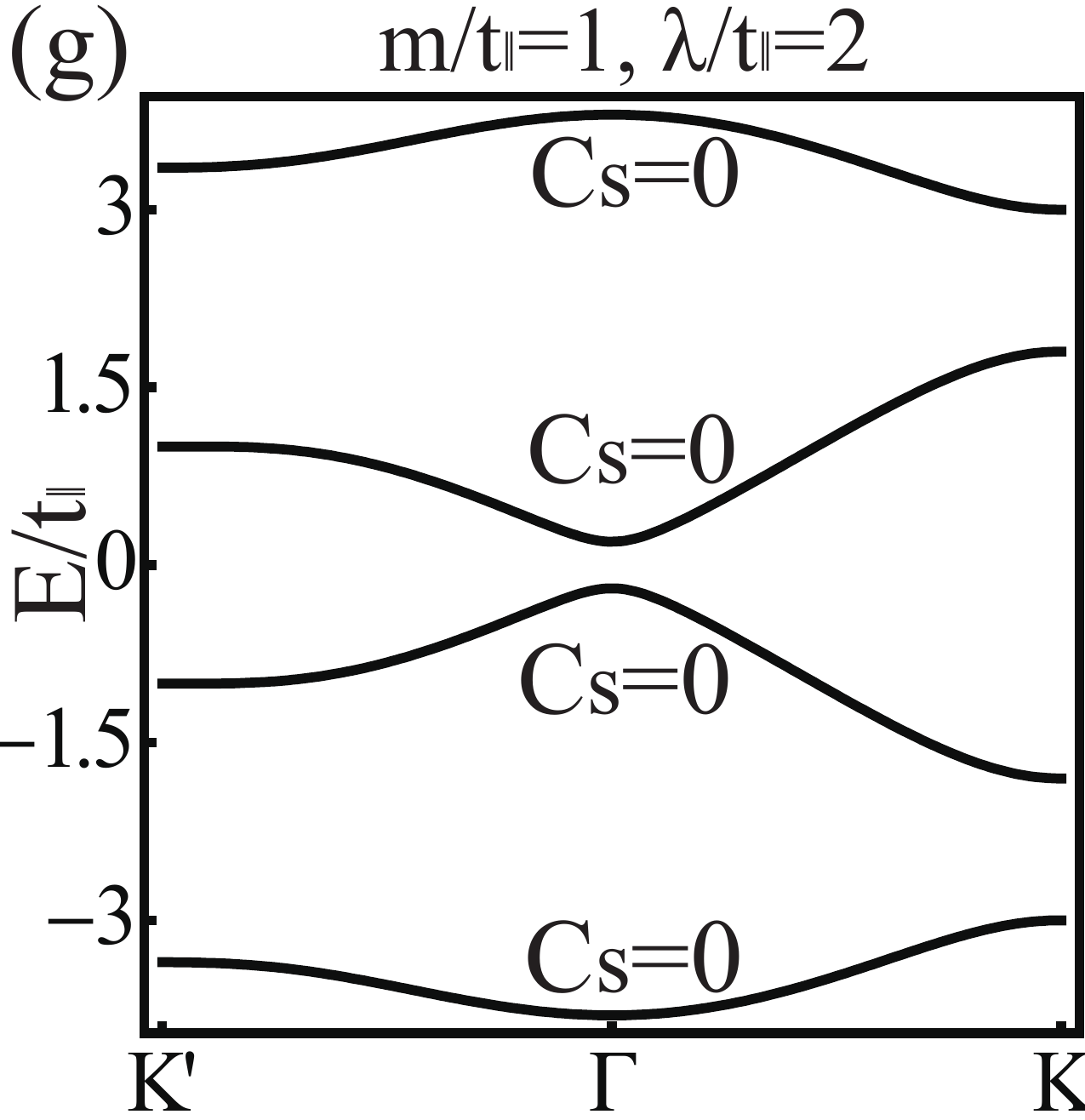}
\caption{The same plot as in Fig. \ref{fig:path1} but for the
evolution fixing $m/t_{\pp}=1$ and increasing $\lambda$
from 0.2 (a) to 2 (g), which passes phases $A_1$, $C_2$, and $B_2$
and $C_1$.
Gaps are closed at $K^\prime$, $K$, and $\Gamma$ points in
(b), (d), and (f), respectively.
}\label{fig:path3}
\end{figure*}

\subsection{Evolution of the topological band structures}
The lattice asymmetry term $m$ by itself can open a gap at $K$ and $K^\prime$
in the absence of SO coupling.
In this case, the gap value is $m$ at both $K$ and $K^\prime$.
The lower two bands remain touched at the $\Gamma$ point with quadratic
band touching.
Nevertheless, the overall band structure remains nontopological.

The SO coupling $\lambda$ brings nontrivial band topology.
Its competition with the lattice asymmetry results in
a rich structure of band structure topology presented
in Fig.~\ref{fig:phase}, which are characterized by
their pattern of spin Chern numbers.
There are two phases characterized by the same spin Chern number
pattern $(1,-1,1-1)$ marked as $A_1$ and $A_2$, respectively;
two phases characterized by $(1,0,0,-1)$ marked as $B_1$ and
$B_2$; and two trivial phases denoted as $C_1$ and $C_2$ $(0,0,0,0)$.

Even an infinitesimal value of $\lambda$ removes the quadratic
band touching between the band 1 and 2, and brings nontrivial band
topology.
The line of $m=0$ corresponds to the situation investigated in the
QAH insulator based on the $p_x$- and $p_y$-orbital bands in the
honeycomb lattice \cite{wu2008,zhang2011a}.
The current situation is a 2D topological insulator with $s_z$
conserved, which is just a double copy of the previous QAH model.
At small values of  $\lambda$, the system is in the $B_1$ phase.
It enters the $A_2$ phase after crossing the line $L_1$
at $\lambda=\frac{3}{2} t_\pp$.

If the system begins with a nonzero lattice asymmetry parameter $m$,
it first enters the $A_1$.
If we increase SO coupling strength $\lambda$ by fixing $m$ at
different values, different band topology transitions appear.
To further clarify these transitions, we plot the spectra
evolutions with increasing $\lambda$ while fixing $m=0.3,0.5$,
and 1 in Figs.~\ref{fig:path1}, ~\ref{fig:path2},
~\ref{fig:path3}, respectively.
Only the spectra along the line cut from $K^\prime$ to $\Gamma$ to
$K$ in the Brillouin zone are plotted.
At small values of $m$ as shown in Fig.~\ref{fig:path1},
the gap first closes at $K^\prime$, and then at $\Gamma$,
and finally at $K$ with increasing $\lambda$.
The sequence of phase transitions is $A_1\rightarrow B_1 \rightarrow
A_2 \rightarrow C_1$.
At intermediate values of $m$ shown in Fig.~\ref{fig:path2},
the gap first closes at $K^\prime$, then at $K$, and finally
at $\Gamma$ leading to a sequence of phase transitions
 $A_1\rightarrow B_1\rightarrow B_2 \rightarrow C_1$.
At large values of $m$ as shown in Fig.~\ref{fig:path3},
the gap first closes at $K^\prime$, then at $K$, and finally at
$\Gamma$.
The sequence of phases is $A_1\rightarrow C_2 \rightarrow
B \rightarrow C_1$.

\section{Reduced two-band models around band crossings}
\label{sect:twoband}

In order to further clarify topological band transitions, we
derive the effective two-band Hamiltonians around the
gap closing points ($\Gamma$, $K$, and $K'$) respectively in this section.

Since the crossing at the $\Gamma$ point occurs at zero energy,
we consider the middle two states.
We construct the two bases as
\bea
|\phi_2(\vec k)\rangle &=& \cos\frac{\alpha}{2}
|\psi_{A,-}(\vec k)\rangle +\sin\frac{\alpha}{2}
|\psi_{B,-}(\vec k)\rangle \nn \\
|\phi_3(\vec k)\rangle &=& -\sin\frac{\alpha}{2}
|\psi_{A,+}(\vec k)\rangle +\cos\frac{\alpha}{2}
|\psi_{B,+}(\vec k)\rangle,
\label{eq:Go}
\eea
where $\alpha=\arctan \frac{3t_\pp}{2m}$.
Right at the $\Gamma$ point, these two bases are the
eigenvectors of the middle two bands with
energies are $E_{2,3}(\Gamma)= \mp(
\sqrt{m^2+(\frac{3}{2}t_\pp)^2}-\lambda)$, respectively.
As $\lambda \rightarrow \sqrt{m^2+ (\frac{3}{2}t_\pp)^2}$,
we construct the low-energy Hamiltonian
for $\vec k$ around the $\Gamma$ point by using
$|\phi_{2,3}(\vec k)\rangle$ as bases:
\bea
&&\left[
\begin{array}{cc}
\avg{\phi_2|H|\phi_2}&\avg{\phi_2|H|\phi_3}\\
\avg{\phi_3|H|\phi_2}&\avg{\phi_3|H|\phi_3}
\end{array}
\right]\nn \\
&=&\left[
\begin{array}{cc}
-\lambda+\sqrt{m^2+(\frac{3}{2}t_\pp)^2}& \frac{3}{4}t_\pp (k_x+ik_y)\\
\frac{3}{4}t_\pp (k_x-ik_y)& \lambda-\sqrt{m^2+(\frac{3}{2}t_\pp)^2}
\end{array}
\right],  \ \ \
\eea
which describes the band crossing of line $L_1$ in Fig.~\ref{fig:phase}.
The two-band effective model for the crossing at the $K^\prime$ point
is just what we have constructed in Eq.~(\ref{eq:biggap}).
It describes the crossing at zero energy represented by
line $L_2$ in Fig.~\ref{fig:phase}.

As for the band crossing at the $K$ point, it occurs between
band 1 and 2, and between 3 and 4 symmetrically with respect
to zero energy ($B_2$, $C_1$, and $C_2$ phases).
For simplicity, we only consider the effective two-band
model at small values of $m$.
In this case, the band crossing is described by line
$L_3$ in Fig.~\ref{fig:phase} occurring at large values of $\lambda\gg m$.
The on-site energy level splitting between the states of
$(p_+,\uparrow)$ and $(p_-,\uparrow)$ is larger than the
hopping integral $t_\pp$, and each of them will develop
a single band in the honeycomb lattice.
The bands of $p_{\pm}$ orbitals lie symmetrically with
respect to zero energy.
Nevertheless, as shown in Refs.~\onlinecite{wu2008,zhang2011a},
the interband coupling at the second-order perturbation level
effectively generates the complex-valued next-nearest-neighbor
hopping as in Haldane's QAH model \cite{haldane1988}.
Our current situation is a TR double copy and thus it
gives rise to the Kane-Mele model.

To describe the above physics, we only keep the $p_{+}$ orbitals
on each site in the case of large values of $\lambda$.
Then the terms of $h_{11}$, $h_{22}$, $h_{21}$, and $h_{12}$
in Eq.~(\ref{eq:HK_up}) become perturbations.
By the second-order perturbation theory, we derive the low-energy
Hamiltonian of $(p_{A,+}(\vec k), p_{B,+}(\vec k))$ bands as
\bea
&&\left[
\begin{array}{cc}
\avg{\psi_{A+}|H|\psi_{A+}}&\avg{\psi_{A+}|H|\psi_{B_+}}\\
\avg{\psi_{B+}|H|\psi_{A+}}&\avg{\psi_{B+}|H|\psi_{B_+}}
\end{array}
\right]\nn \\
&=&\left[
\begin{array}{cc}
m+m_H(\vec k) &-\frac{t_\pp}{2}l^*(\vec k)\nn \\
-\frac{t_\pp}{2}l(\vec k)& -m-m_H(\vec k)
\end{array}
\right],
\eea
where
\bea
m_H(\vec k)=\frac{\sqrt{3}}{8}\frac{t^2_\pp}{\lambda}
\sin \eta_s(\vec k).
\eea
Around the $K$ point, $m_H(K)=-\frac{9}{16}\frac{t_\pp^2}{\lambda}$.
The band crossing occurs when $m+m_H(\vec k)$ switches sign,
which gives rise to line $L_3$ in Fig. \ref{fig:phase}.

The topological gap opens at the $K^\prime$ point between bands 2 and 3.
According to Eq.~(\ref{eq:newham}), we only need to keep
the right-bottom block for the construction of the low-energy
two-band model.
By expanding around the $K^\prime$ point, we have
\bea
&&\left[
\begin{array}{cc}
\avg{A_2|H|A_2}&\avg{A_2|H|B_2}\\
\avg{B_2|H|A_2}&\avg{B_2|H|B_2}
\end{array}
\right]\nn \\
&=&\left[
\begin{array}{cc}
m-\lambda& -\frac{3}{4}t_\pp (\delta k_x+i \delta k_y)\\
-\frac{3}{4}t_\pp (\delta k_x-i\delta k_y)& -m+\lambda
\end{array}
\right],
\label{eq:biggap}
\eea
where $\delta \vec k= \vec k-\vec K^\prime$, and thus the mass term is controlled
by $m-\lambda$. For completeness, we also derive the effective two-band Hamiltonian for bands 2 and 3 around the $K$ point similarly,
which yields the gap value $m+\lambda$.
In the absence of lattice asymmetry, the gap values at
$K$ and $K^\prime$ are both the SO coupling strength.

Now let us look more carefully at the eigen wave functions of
the effective two-band Hamiltonian for bands 2 and 3 at
$K^\prime$ and $K$ points and check their orbital angular momenta.
The eigenstates are just $|A_2(K^\prime)\rangle$ and
$|B_2(K^\prime)\rangle$ at $K^\prime$, and
$|A_2(K)\rangle$, and $|B_2(K)\rangle$ at $K$.
In the bases of Eq.~(\ref{eq:bases1}), we express
\bea
&&|A_2(K^\prime)\rangle=\left[
\begin{array}{c}
1\\
0\\
0\\
0
\end{array}
\right], \ \ \,
|B_2(K^\prime)\rangle=\left[
\begin{array}{c}
0\\
0\\
0\\
1
\end{array}
\right],\nn \\
&&
|A_2(K)\rangle=\left[
\begin{array}{c}
0\\
0\\
1\\
0
\end{array}
\right], \ \ \,
|B_2(K)\rangle=\left[
\begin{array}{c}
0\\
1\\
0\\
0
\end{array}
\right].
\label{eq:Kpo}
\eea
All of them are the orbital angular momentum eigenstates
with $L_z=\pm 1$.
Considering this is the sector with $s=\uparrow$,
the gap is just the atomic SO coupling strength
$\lambda$ in the absence of the lattice asymmetry term
$m$.

\section{Large topological band gaps}
\label{sect:largegap}

\begin{figure*}
\centering\includegraphics[width=.8\linewidth]{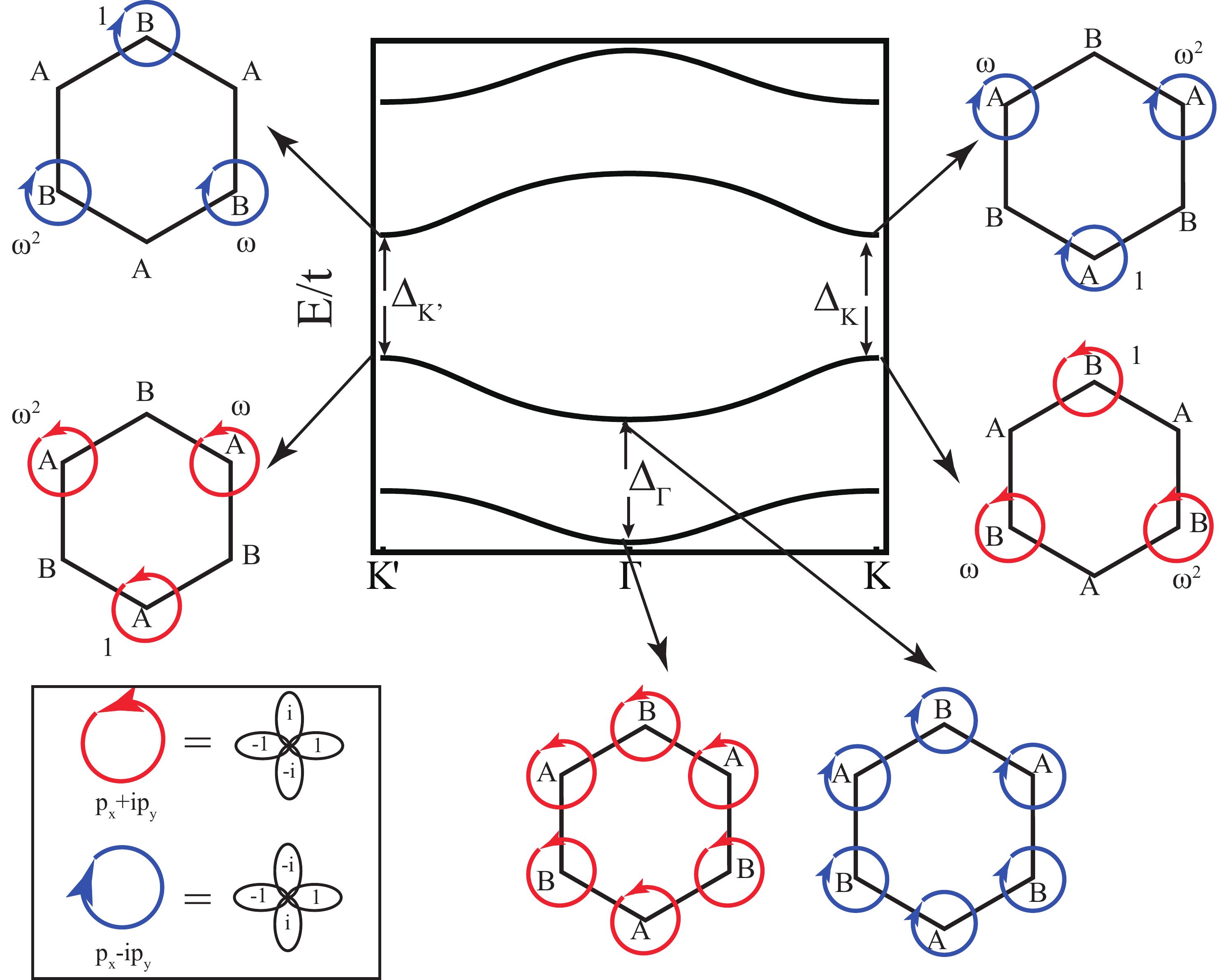}
\caption{(Color online) The topological energy gap at the high symmetry $K'$, $K$, and $\Gamma$ points for the $B_1$
  phase, denoted $\Delta_{K'}$, $\Delta_{K}$, and $\Delta_{\Gamma}$
  respectively. $\Delta_{K'}=\Delta_{K}=\Delta_{\Gamma}=2\lambda$ in the absence of lattice asymmetry. The corresponding
  real space orbital configurations of the eigenstates are of $p_x \pm ip_y$ type ($L_z=\pm 1$). We denote the orbital
  angular momentum $L_z=\pm 1$ eigenstates as red (blue) counterclockwise (clockwise) circles, as shown in the inset. At
  the $K'(K)$ point, the eigenstate for band 2 is of the $p_x+ip_y$ type completely located at the A(B) sublattice,
  while for band 3 the eigenstate is of the $p_x-ip_y$ type at the B(A) sublattice. The phase factor for each site in
  the hexagonal plquette is $\omega=e^{i \frac{2}{3} \pi}$. At the $\Gamma$ point, the eigenstates for band 1 and 2 are
  both superpositions of the wave functions at $A$ and $B$ sublattices, but with opposite orbital angular momentum.  }
\label{fig:gap}
\end{figure*}

The most striking feature of the these $p_x$-$p_y$ systems is the large
topological band gap at $K'$, $K$, and $\Gamma$ points.
In this section, we analyze the origin of large topological band
gaps at these $k$ points in the $B_1$ phase (quantum spin Hall (QSH) phase, $\lambda>m$).
For the case of a single-component fermion QAH model studied in
Ref.~\onlinecite{wu2008}, it has been analyzed that the gap values
at the $\Gamma$, $K$, and $K^\prime$ points
are just the on-site rotation angular velocity $\Omega$
in the absence of the lattice asymmetry term.
The situation in this paper is a TR invariant double copy of the
previously single component case, and thus the role of
of $\Omega$ is replaced by the on-site atomic SO coupling strength
$\lambda$.

At the $K'$ point, according to Eq.~(\ref{eq:Kpo}), the eigenstates
for the bands 2,3 are orbital angular momentum eigenstates with $L_z=\pm 1$.
The energy and corresponding eigenstates for bands 2 and 3 are
\bea
E_2(K')&=& m-\lambda, \quad |\phi_2(K')\rangle = |\psi_{A,+}(K')\rangle, \nn \\
E_3(K')&=& \lambda-m, \quad |\phi_3(K')\rangle = |\psi_{B,-}(K')\rangle, \nn \\
\Delta_{K'}&=& 2 (\lambda - m).
\eea
As shown in Fig.~\ref{fig:gap}, the eigenstate for band 2 has $L_z=+1$
with the energy $m-\lambda$, which is of $p_x+i p_y$ type, and
its wave function is totally on the $A$ sublattice.
In contrast, the eigenstate for band 3 has $L_z=-1$
with the energy $\lambda-m$.
It is of the $p_x-i p_y$ type whose wave function completely distributes on
the $B$ sublattice.
The topological band gap is thus $2(\lambda-m)$.
If the sublattice asymmetry term vanishes, i.e., $m=0$,
the band gap is just $2\lambda$.

Obviously, the atomic on-site SO coupling strength $\lambda$ directly
contributes to the topological band gap, leading to a large band splitting.
It is because at the $K'$ point, the eigenstates of the system are also $L_z$
eigenstates, which means the topological band gap is the eigenenergy
difference between the SO coupling term $s_z L_z$ for $L_z=\pm 1$.
It is easy to generalize the analysis to the $K$ point similarly.

At the $\Gamma$ point, the Hamiltonian $H(\vec k)$ preserves all
the rotation symmetries of the system, and thus
the SO coupling term $s_z L_z$ commutes with $H(\vec k)$.
The eigenstates simultaneously diagonalize the SO coupling term
and $H(\vec k)$.
The energy and corresponding eigenstates for bands 1 and 2 at
the $\Gamma$ point are
\bea
E_1(\Gamma) &=& -\lambda-\sqrt{m^2+(\frac{3}{2}t_{\pp})^2}, \nn \\
|\phi_1(\Gamma)\rangle &=& \sin\frac{\alpha}{2}
|\psi_{A,+}(\Gamma)\rangle +\cos\frac{\alpha}{2}
|\psi_{B,+}(\Gamma)\rangle, \nn \\
E_2(\Gamma)&=& \lambda-\sqrt{m^2+(\frac{3}{2}t_{\pp})^2}, \nn \\
 |\phi_2(\Gamma)\rangle &=& \cos\frac{\alpha}{2}
|\psi_{A,-}(\Gamma)\rangle +\sin\frac{\alpha}{2}
|\psi_{B,-}(\Gamma)\rangle, \nn \\
\Delta_{\Gamma}&=& 2\lambda.
\eea
The eigenstates for bands 1,2 are the superpositions of wave functions on
both the $A$ and $B$ sublattices.
However, for band 1, the eigenstate is an $L_z=-1$ eigenstate, and the
eigenstate for band 2 is an $L_z=1$ eigenstate (see Fig.~\ref{fig:gap}).
As a result, the topological band gap $\Delta_{\Gamma}$ is the
energy difference of the SO coupling term $s_z L_z$, which is $2\lambda$.

We discuss the dependence of the topological gap values on SO coupling strength in the $B_1$ phase. Let us first
consider the gap between the lowest two bands. For the case without lattice asymmetry, i.e., $m/t_{\pp}=0$, in the weak
and intermediate regimes of the SO coupling strength $0< \lambda/t_{\pp}<3/(4\sqrt{2})$, the minimal gap is located at
the $\Gamma$ point as shown in Fig.~\ref{fig:path1}(b).  In typical solid state systems, $\lambda$ lies in these
regimes, and thus, typically, the topological gap can approach up to $2\lambda=3/(2\sqrt{2}) t_{\pp}$, which is a very
large gap.  If $\lambda$ further increases, then the minimal gap shifts from the $\Gamma$ point to the $K$ points, and
the value of the gap shrinks as $\lambda$ increases. Similarly, consider the topological gap between the middle two
bands, and for parameters $m/t_{\pp}=0$: as long as the SO coupling strength is in the $B_1$ phase, the minimal gap is
located at the $K(K')$ point, which can approach up to $2\lambda=3\, t_{\pp}$.

\section{Quantum Anomalous Hall State}
\label{sect:QAHE}

In this section, we add the N\'eel antiferromagnetic exchange field
term [Eq.~(\ref{eq:neel})] to the Hamiltonian.
This term gives rise to another mass generation mechanism.
Together with the atomic SO coupling term of $L_z \sigma_z$, and
the sublattice asymmetry term [Eq.~(\ref{eq:m})] discussed before,
we can drive the system to a QAH state.
A similar mechanism was also presented in the single-orbital honeycomb
lattice \cite{Liang2013a}, and here we generalize it
to the $p_x$-$p_y$-orbital systems.

We consider the gap opening at the $K$ and $K^\prime$ points,
and assume that bands 1 and 2 are filled.
In the absence of the N\'eel term [Eq.~(\ref{eq:neel})], the system
is in the trivially gapped phase  $A_1$ at $m> \lambda $,
and in the QSH phase $B_1$ at $\lambda> m$.

\begin{figure}
\centering\includegraphics[width=0.5\linewidth]{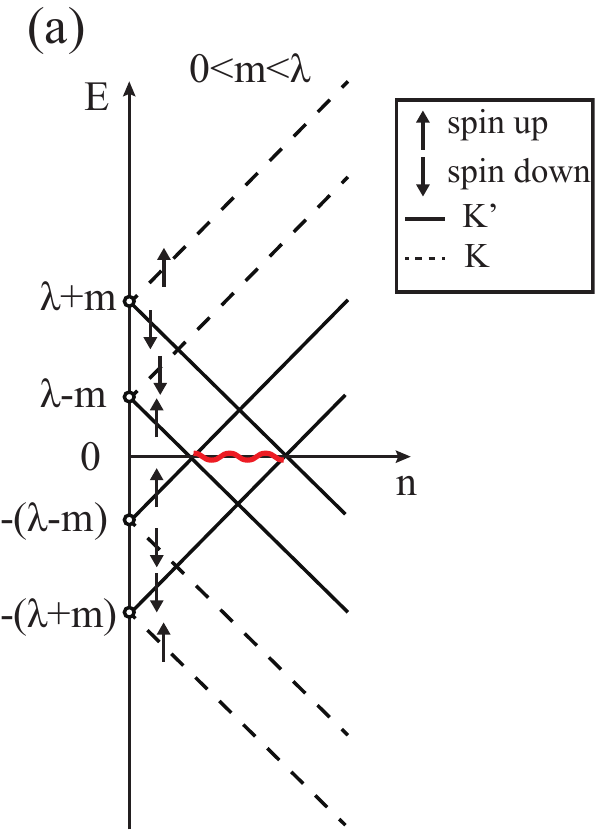}
\centering\includegraphics[width=0.36\linewidth]{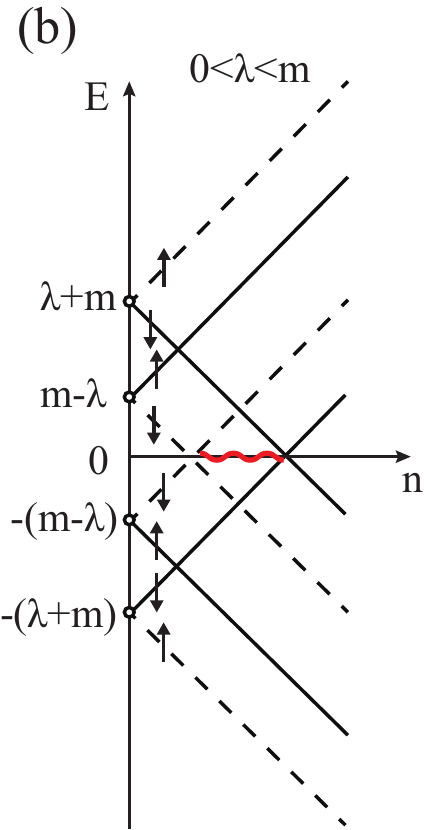}
\caption{(Color online)
Diagrams of the energy level crossing with increasing
antiferromagnetic exchange field strength $n$ for two parameter regimes
(a) $0<m<\lambda$ and (b) $0<\lambda<m$.
Red wavy lines indicate the range of $n$ for the
system to be in a QAH phase ($|\lambda-m|<n<\lambda+m$).
}\label{fig:QAHE}
\end{figure}

Let us start with the QSH phase $B_1$ with $\lambda>m>0$, and
gradually turn on the N\'eel exchange magnitude $n>0$.
The energy levels for different spin sectors at the $K'$ and $K$ points
for the middle two bands are
\bea
E_{2,3, \uparrow}(K')&=& \mp (\lambda-m-n), \nn \\
E_{2,3, \downarrow}(K')&=& \mp (\lambda+m-n), \nn \\
E_{2,3,\uparrow}(K)&=& \mp(\lambda+m+n),\nn \\
E_{2,3,\downarrow}(K)&=& \mp (\lambda-m+n).
\eea
The gap will not close for both spin-$\uparrow$ and spin-$\downarrow$ sectors at the
$K$ point with increasing $n$, and thus we focus on the
band crossing at the $K^\prime$ point.
At this point, the first band crossing occurs in the spin-$\uparrow$ sector
at $n=\lambda-m$, which changes the spin-$\uparrow$ sector into the topologically
trivial regime.
Meanwhile, the spin-$\downarrow$ sector remains topologically nontrivial,
and thus the system becomes a QAH state.
If we further increase $n$, the second band crossing occurs in the
spin-$\downarrow$ sector at $n=\lambda+m$, at which the spin-$\downarrow$ sector
also becomes topologically trivial.
In this case, the entire system is a trivial band insulator.
The QAH state can be realized for $\lambda-m<n<\lambda+m$.
The band crossing diagrams are shown in Fig.~\ref{fig:QAHE}(a).

Similarly, we start from the $A_1$ trivially gapped phase ($0<\lambda<m$),
and gradually turn on the N\'eel exchange field $n$.
The middle two energy levels for both spin sectors at the $K'$ and $K$ points are
\bea
E_{2,3, \uparrow}(K')&=& \mp( m-\lambda+n),  \nn \\
E_{2,3,\downarrow}(K')&=& \mp(m+\lambda-n) , \nn \\
E_{2,3,\uparrow}(K)&=& \mp(m+ \lambda+n) ,\nn \\
E_{2,3,\downarrow}(K)&=&  \mp(m-\lambda-n).
\eea
In this case, the spin-$\uparrow$ sector remains in the trivially gapped
phase with increasing $n$, since there is no band inversion in
this sector [see Fig.~\ref{fig:QAHE}(b)].
The first band crossing occurs in the spin-$\downarrow$ sector at the $K$ point
when $n=m-\lambda$, rendering this sector topologically nontrivial,
and then the whole system goes into a QAH phase.
The second band inversion occurs at the $K'$ point also in
the spin-$\downarrow$ sector at $n=\lambda+m$.
Now the spin-$\downarrow$ sector is back into a topologically trivial phase,
and the whole system is a trivial band insulator for $n>\lambda+m$.
Similarly to the previous case, the QAH phase is realized
at $-\lambda+m<n<\lambda+m$.

There are three gap parameters in our model, the spin-orbit coupling $\lambda$, the sublattice asymmetry term $m$, and
the N\'eel exchange field $n$. Combining the two situations discussed above, we summarize the condition for the
appearance of the QAH state as follows:
\begin{eqnarray}
|\lambda-m|<n<\lambda+m,
\end{eqnarray}
which is also equivalent to $|m-n|<\lambda<m+n$, or $|\lambda-n|<m<\lambda+n$.  In other words, the three gap parameters
$\lambda$, $m$, and $n$ can form a triangle.  For the buckled honeycomb lattices, the A and B sublattices are at
different heights.  The N\'eel exchange field $n$ can be generated by attaching two ferromagnetic substrates with
opposite magnetizations to the two surfaces, and the sublattice asymmetry term $m$ can also be generated if the contacts
with these two substrates are asymmetric.  In the parameter regime for the QAH state, it is easy to check that the
maximal topological gap is the minimum of $\lambda$ and $m$.

\section{Conclusions and Outlooks}
\label{sect:conclusion}
In summary, we have presented a minimal model to describe the 2D topological insulator states in the honeycomb lattice
which have been recently proposed in the literature.  The $p_x$ and $p_y$ orbitals are the key, and thus their
properties are dramatically different from those in graphene.  The atomic level SO coupling directly contributes to the
topological gap opening, and thus the gap can be large.  Due to the conservation of $s_z$, the band structures are a TR
invariant doublet of the previously investigated QAHE based on the $p$ orbital in the honeycomb lattice.  The band
topology is described by the spin Chern numbers.  Both sublattice asymmetry and the on-site SO coupling can open the
gap, and their competition leads to a rich structure of topological band insulating phases.  Due to the underlying
structure of Clifford algebra, the energy spectra and eigen wave functions can be obtained analytically.  Also, the
transition lines among different topological insulators are also analytically obtained.  Low-energy two-band models are
constructed around band crossings.  Furthermore, with the help of the N\'eel antiferromagnetic exchange field, the model
can enter into a QAH phase.  This work provides a useful platform for further exploring interaction and topological
properties in such systems.

In addition to a class of solid state materials, the model constructed in this article can, in principle, also be
realized in ultracold-atom optical lattices. For example, in previous papers by one of the authors and his collaborators
(Refs. \onlinecite{wu2008b} and \onlinecite{zhang2011a}), the quantum anomalous Hall models were proposed for spinless
fermions of the $p_x/p_y$ bands in the honeycomb optical lattices. By this technique, each optical site is rotating
around its own center, which can be modeled as an orbital Zeeman term. The quantum spin Hall model of
Eqs. (\ref{eq:ham0}) and (\ref{eq:so}) is a time-reversal invariant double of the anomalous quantum Hall model, which in
principle can be realized by the spin-dependent on-site rotations of the honeycomb lattice, i.e., the rotation angular
velocities for spin-$\uparrow$ and spin-$\downarrow$ fermions are opposite to each other. This is essentially a
spin-orbit coupling term $L_z\, S_z$ and the rotation angular velocity plays the role of spin-orbit coupling
strength. In order to observe the topological phase, we need the fermions population to fill the $p$-orbital
bands. Then, the phase diagram will be the same as in Fig.\ref{fig:phase}, by replacing the spin-orbit coupling strength
with the magnitude of the angular velocity.

{\it Note added} Near the completion of this work, we became aware of the work Ref.~\onlinecite{liu2014} in which the
low energy effective model of the 2D topological insulators on honeycomb lattice are also constructed.

\begin{acknowledgments}
  G. F. Z. and C. W. are supported by the NSF DMR-1410375 and AFOSR FA9550-11-1-0067(YIP).  Y. L. thanks the Inamori
  Fellowship and the support at the Princeton Center for Theoretical Science.  C.W. acknowledges financial support from
  the National Natural Science Foundation of China (11328403).
\end{acknowledgments}


\begin{thebibliography}{47}%
\makeatletter
\providecommand \@ifxundefined [1]{%
 \@ifx{#1\undefined}
}%
\providecommand \@ifnum [1]{%
 \ifnum #1\expandafter \@firstoftwo
 \else \expandafter \@secondoftwo
 \fi
}%
\providecommand \@ifx [1]{%
 \ifx #1\expandafter \@firstoftwo
 \else \expandafter \@secondoftwo
 \fi
}%
\providecommand \natexlab [1]{#1}%
\providecommand \enquote  [1]{``#1''}%
\providecommand \bibnamefont  [1]{#1}%
\providecommand \bibfnamefont [1]{#1}%
\providecommand \citenamefont [1]{#1}%
\providecommand \href@noop [0]{\@secondoftwo}%
\providecommand \href [0]{\begingroup \@sanitize@url \@href}%
\providecommand \@href[1]{\@@startlink{#1}\@@href}%
\providecommand \@@href[1]{\endgroup#1\@@endlink}%
\providecommand \@sanitize@url [0]{\catcode `\\12\catcode `\$12\catcode
  `\&12\catcode `\#12\catcode `\^12\catcode `\_12\catcode `\%12\relax}%
\providecommand \@@startlink[1]{}%
\providecommand \@@endlink[0]{}%
\providecommand \url  [0]{\begingroup\@sanitize@url \@url }%
\providecommand \@url [1]{\endgroup\@href {#1}{\urlprefix }}%
\providecommand \urlprefix  [0]{URL }%
\providecommand \Eprint [0]{\href }%
\providecommand \doibase [0]{http://dx.doi.org/}%
\providecommand \selectlanguage [0]{\@gobble}%
\providecommand \bibinfo  [0]{\@secondoftwo}%
\providecommand \bibfield  [0]{\@secondoftwo}%
\providecommand \translation [1]{[#1]}%
\providecommand \BibitemOpen [0]{}%
\providecommand \bibitemStop [0]{}%
\providecommand \bibitemNoStop [0]{.\EOS\space}%
\providecommand \EOS [0]{\spacefactor3000\relax}%
\providecommand \BibitemShut  [1]{\csname bibitem#1\endcsname}%
\let\auto@bib@innerbib\@empty
\bibitem [{\citenamefont {Klitzing}\ \emph {et~al.}(1980)\citenamefont
  {Klitzing}, \citenamefont {Dorda},\ and\ \citenamefont
  {Pepper}}]{klitzing1980}%
  \BibitemOpen
  \bibfield  {author} {\bibinfo {author} {\bibfnamefont {K.}~\bibnamefont
  {Klitzing}}, \bibinfo {author} {\bibfnamefont {G.}~\bibnamefont {Dorda}}, \
  and\ \bibinfo {author} {\bibfnamefont {M.}~\bibnamefont {Pepper}},\ }\href
  {\doibase 10.1103/PhysRevLett.45.494} {\bibfield  {journal} {\bibinfo
  {journal} {Phys. Rev. Lett.}\ }\textbf {\bibinfo {volume} {45}},\ \bibinfo
  {pages} {494} (\bibinfo {year} {1980})}\BibitemShut {NoStop}%
\bibitem [{\citenamefont {Thouless}\ \emph {et~al.}(1982)\citenamefont
  {Thouless}, \citenamefont {Kohmoto}, \citenamefont {Nightingale},\ and\
  \citenamefont {den Nijs}}]{thouless1982}%
  \BibitemOpen
  \bibfield  {author} {\bibinfo {author} {\bibfnamefont {D.}~\bibnamefont
  {Thouless}}, \bibinfo {author} {\bibfnamefont {M.}~\bibnamefont {Kohmoto}},
  \bibinfo {author} {\bibfnamefont {M.}~\bibnamefont {Nightingale}}, \ and\
  \bibinfo {author} {\bibfnamefont {M.}~\bibnamefont {den Nijs}},\ }\href
  {\doibase 10.1103/PhysRevLett.49.405} {\bibfield  {journal} {\bibinfo
  {journal} {Phys. Rev. Lett.}\ }\textbf {\bibinfo {volume} {49}},\ \bibinfo
  {pages} {405} (\bibinfo {year} {1982})}\BibitemShut {NoStop}%
\bibitem [{\citenamefont {Halperin}(1982)}]{halperin1982}%
  \BibitemOpen
  \bibfield  {author} {\bibinfo {author} {\bibfnamefont {B.}~\bibnamefont
  {Halperin}},\ }\href {\doibase 10.1103/PhysRevB.25.2185} {\bibfield
  {journal} {\bibinfo  {journal} {Phys. Rev. B}\ }\textbf {\bibinfo {volume}
  {25}},\ \bibinfo {pages} {2185} (\bibinfo {year} {1982})}\BibitemShut
  {NoStop}%
\bibitem [{\citenamefont {Kohmoto}(1985)}]{kohmoto1985}%
  \BibitemOpen
  \bibfield  {author} {\bibinfo {author} {\bibfnamefont {M.}~\bibnamefont
  {Kohmoto}},\ }\href {\doibase 10.1016/0003-4916(85)90148-4} {\bibfield
  {journal} {\bibinfo  {journal} {Ann. Phys. (N. Y).}\ }\textbf {\bibinfo
  {volume} {160}},\ \bibinfo {pages} {343} (\bibinfo {year}
  {1985})}\BibitemShut {NoStop}%
\bibitem [{\citenamefont {Haldane}(1988)}]{haldane1988}%
  \BibitemOpen
  \bibfield  {author} {\bibinfo {author} {\bibfnamefont {F.~D.~M.}\
  \bibnamefont {Haldane}},\ }\href {\doibase 10.1103/PhysRevLett.61.2015}
  {\bibfield  {journal} {\bibinfo  {journal} {Phys. Rev. Lett.}\ }\textbf
  {\bibinfo {volume} {61}},\ \bibinfo {pages} {2015} (\bibinfo {year}
  {1988})}\BibitemShut {NoStop}%
\bibitem [{\citenamefont {Qi}\ and\ \citenamefont {Zhang}(2010)}]{qi2010a}%
  \BibitemOpen
  \bibfield  {author} {\bibinfo {author} {\bibfnamefont {X.~L.}\ \bibnamefont
  {Qi}}\ and\ \bibinfo {author} {\bibfnamefont {S.~C.}\ \bibnamefont {Zhang}},\
  }\href {\doibase 10.1063/1.3293411} {\bibfield  {journal} {\bibinfo
  {journal} {Phys. Today}\ }\textbf {\bibinfo {volume} {63}},\ \bibinfo {pages}
  {33} (\bibinfo {year} {2010})}\BibitemShut {NoStop}%
\bibitem [{\citenamefont {Qi}\ and\ \citenamefont {Zhang}(2011)}]{qi2011}%
  \BibitemOpen
  \bibfield  {author} {\bibinfo {author} {\bibfnamefont {X.~L.}\ \bibnamefont
  {Qi}}\ and\ \bibinfo {author} {\bibfnamefont {S.~C.}\ \bibnamefont {Zhang}},\
  }\href {\doibase 10.1103/RevModPhys.83.1057} {\bibfield  {journal} {\bibinfo
  {journal} {Rev. Mod. Phys.}\ }\textbf {\bibinfo {volume} {83}},\ \bibinfo
  {pages} {1057} (\bibinfo {year} {2011})}\BibitemShut {NoStop}%
\bibitem [{\citenamefont {Hasan}\ and\ \citenamefont {Kane}(2010)}]{Hasan2010}%
  \BibitemOpen
  \bibfield  {author} {\bibinfo {author} {\bibfnamefont {M.~Z.}\ \bibnamefont
  {Hasan}}\ and\ \bibinfo {author} {\bibfnamefont {C.~L.}\ \bibnamefont
  {Kane}},\ }\href {\doibase 10.1103/RevModPhys.82.3045} {\bibfield  {journal}
  {\bibinfo  {journal} {Rev. Mod. Phys.}\ }\textbf {\bibinfo {volume} {82}},\
  \bibinfo {pages} {3045} (\bibinfo {year} {2010})}\BibitemShut {NoStop}%
\bibitem [{\citenamefont {Kane}\ and\ \citenamefont
  {Mele}(2005{\natexlab{a}})}]{kane2005}%
  \BibitemOpen
  \bibfield  {author} {\bibinfo {author} {\bibfnamefont {C.~L.}\ \bibnamefont
  {Kane}}\ and\ \bibinfo {author} {\bibfnamefont {E.~J.}\ \bibnamefont
  {Mele}},\ }\href {\doibase 10.1103/PhysRevLett.95.226801} {\bibfield
  {journal} {\bibinfo  {journal} {Phys. Rev. Lett.}\ }\textbf {\bibinfo
  {volume} {95}},\ \bibinfo {pages} {226801} (\bibinfo {year}
  {2005}{\natexlab{a}})}\BibitemShut {NoStop}%
\bibitem [{\citenamefont {Bernevig}\ and\ \citenamefont
  {Zhang}(2006)}]{bernevig2006}%
  \BibitemOpen
  \bibfield  {author} {\bibinfo {author} {\bibfnamefont {B.~A.}\ \bibnamefont
  {Bernevig}}\ and\ \bibinfo {author} {\bibfnamefont {S.~C.}\ \bibnamefont
  {Zhang}},\ }\href {\doibase 10.1103/PhysRevLett.96.106802} {\bibfield
  {journal} {\bibinfo  {journal} {Phys. Rev. Lett.}\ }\textbf {\bibinfo
  {volume} {96}},\ \bibinfo {pages} {106802} (\bibinfo {year}
  {2006})}\BibitemShut {NoStop}%
\bibitem [{\citenamefont {Bernevig}\ \emph {et~al.}(2006)\citenamefont
  {Bernevig}, \citenamefont {Hughes},\ and\ \citenamefont
  {Zhang}}]{bernevig2006a}%
  \BibitemOpen
  \bibfield  {author} {\bibinfo {author} {\bibfnamefont {B.~A.}\ \bibnamefont
  {Bernevig}}, \bibinfo {author} {\bibfnamefont {T.~L.}\ \bibnamefont
  {Hughes}}, \ and\ \bibinfo {author} {\bibfnamefont {S.~C.}\ \bibnamefont
  {Zhang}},\ }\href {\doibase 10.1126/science.1133734} {\bibfield  {journal}
  {\bibinfo  {journal} {Science}\ }\textbf {\bibinfo {volume} {314}},\ \bibinfo
  {pages} {1757} (\bibinfo {year} {2006})}\BibitemShut {NoStop}%
\bibitem [{\citenamefont {Qi}\ \emph {et~al.}(2008)\citenamefont {Qi},
  \citenamefont {Hughes},\ and\ \citenamefont {Zhang}}]{qi2008a}%
  \BibitemOpen
  \bibfield  {author} {\bibinfo {author} {\bibfnamefont {X.~L.}\ \bibnamefont
  {Qi}}, \bibinfo {author} {\bibfnamefont {T.~L.}\ \bibnamefont {Hughes}}, \
  and\ \bibinfo {author} {\bibfnamefont {S.~C.}\ \bibnamefont {Zhang}},\ }\href
  {\doibase 10.1103/PhysRevB.78.195424} {\bibfield  {journal} {\bibinfo
  {journal} {Phys. Rev. B}\ }\textbf {\bibinfo {volume} {78}},\ \bibinfo
  {pages} {195424} (\bibinfo {year} {2008})}\BibitemShut {NoStop}%
\bibitem [{\citenamefont {Fu}\ and\ \citenamefont {Kane}(2007)}]{fu2007a}%
  \BibitemOpen
  \bibfield  {author} {\bibinfo {author} {\bibfnamefont {L.}~\bibnamefont
  {Fu}}\ and\ \bibinfo {author} {\bibfnamefont {C.~L.}\ \bibnamefont {Kane}},\
  }\href {\doibase 10.1103/PhysRevB.76.045302} {\bibfield  {journal} {\bibinfo
  {journal} {Phys. Rev. B}\ }\textbf {\bibinfo {volume} {76}},\ \bibinfo
  {pages} {045302} (\bibinfo {year} {2007})}\BibitemShut {NoStop}%
\bibitem [{\citenamefont {Moore}\ and\ \citenamefont
  {Balents}(2007)}]{moore2007}%
  \BibitemOpen
  \bibfield  {author} {\bibinfo {author} {\bibfnamefont {J.~E.}\ \bibnamefont
  {Moore}}\ and\ \bibinfo {author} {\bibfnamefont {L.}~\bibnamefont
  {Balents}},\ }\href {\doibase 10.1103/PhysRevB.75.121306} {\bibfield
  {journal} {\bibinfo  {journal} {Phys. Rev. B}\ }\textbf {\bibinfo {volume}
  {75}},\ \bibinfo {pages} {121306(R)} (\bibinfo {year} {2007})}\BibitemShut
  {NoStop}%
\bibitem [{\citenamefont {Roy}(2010)}]{roy2010}%
  \BibitemOpen
  \bibfield  {author} {\bibinfo {author} {\bibfnamefont {R.}~\bibnamefont
  {Roy}},\ }\href {\doibase 10.1088/1367-2630/12/6/065009} {\bibfield
  {journal} {\bibinfo  {journal} {New J. Phys.}\ }\textbf {\bibinfo {volume}
  {12}},\ \bibinfo {pages} {065009} (\bibinfo {year} {2010})}\BibitemShut
  {NoStop}%
\bibitem [{\citenamefont {Zhang}\ and\ \citenamefont {Hu}(2001)}]{zhang2001}%
  \BibitemOpen
  \bibfield  {author} {\bibinfo {author} {\bibfnamefont {S.~C.}\ \bibnamefont
  {Zhang}}\ and\ \bibinfo {author} {\bibfnamefont {J.~P.}\ \bibnamefont {Hu}},\
  }\href {\doibase 10.1126/science.294.5543.823} {\bibfield  {journal}
  {\bibinfo  {journal} {Science}\ }\textbf {\bibinfo {volume} {294}},\ \bibinfo
  {pages} {823} (\bibinfo {year} {2001})}\BibitemShut {NoStop}%
\bibitem [{\citenamefont {K\"{o}nig}\ \emph {et~al.}(2007)\citenamefont
  {K\"{o}nig}, \citenamefont {Wiedmann}, \citenamefont {Br\"{u}ne},
  \citenamefont {Roth}, \citenamefont {Buhmann}, \citenamefont {Molenkamp},
  \citenamefont {Qi},\ and\ \citenamefont {Zhang}}]{konig2007}%
  \BibitemOpen
  \bibfield  {author} {\bibinfo {author} {\bibfnamefont {M.}~\bibnamefont
  {K\"{o}nig}}, \bibinfo {author} {\bibfnamefont {S.}~\bibnamefont {Wiedmann}},
  \bibinfo {author} {\bibfnamefont {C.}~\bibnamefont {Br\"{u}ne}}, \bibinfo
  {author} {\bibfnamefont {A.}~\bibnamefont {Roth}}, \bibinfo {author}
  {\bibfnamefont {H.}~\bibnamefont {Buhmann}}, \bibinfo {author} {\bibfnamefont
  {L.~W.}\ \bibnamefont {Molenkamp}}, \bibinfo {author} {\bibfnamefont {X.-L.}\
  \bibnamefont {Qi}}, \ and\ \bibinfo {author} {\bibfnamefont {S.-C.}\
  \bibnamefont {Zhang}},\ }\href {\doibase 10.1126/science.1148047} {\bibfield
  {journal} {\bibinfo  {journal} {Science}\ }\textbf {\bibinfo {volume}
  {318}},\ \bibinfo {pages} {766} (\bibinfo {year} {2007})}\BibitemShut
  {NoStop}%
\bibitem [{\citenamefont {Hsieh}\ \emph {et~al.}(2008)\citenamefont {Hsieh},
  \citenamefont {Qian}, \citenamefont {Wray}, \citenamefont {Xia},
  \citenamefont {Hor}, \citenamefont {Cava},\ and\ \citenamefont
  {Hasan}}]{hsieh2008}%
  \BibitemOpen
  \bibfield  {author} {\bibinfo {author} {\bibfnamefont {D.}~\bibnamefont
  {Hsieh}}, \bibinfo {author} {\bibfnamefont {D.}~\bibnamefont {Qian}},
  \bibinfo {author} {\bibfnamefont {L.}~\bibnamefont {Wray}}, \bibinfo {author}
  {\bibfnamefont {Y.}~\bibnamefont {Xia}}, \bibinfo {author} {\bibfnamefont
  {Y.~S.}\ \bibnamefont {Hor}}, \bibinfo {author} {\bibfnamefont {R.~J.}\
  \bibnamefont {Cava}}, \ and\ \bibinfo {author} {\bibfnamefont {M.~Z.}\
  \bibnamefont {Hasan}},\ }\href {\doibase 10.1038/nature06843} {\bibfield
  {journal} {\bibinfo  {journal} {Nature}\ }\textbf {\bibinfo {volume} {452}},\
  \bibinfo {pages} {970} (\bibinfo {year} {2008})}\BibitemShut {NoStop}%
\bibitem [{\citenamefont {Zhang}\ \emph {et~al.}(2009)\citenamefont {Zhang},
  \citenamefont {Liu}, \citenamefont {Qi}, \citenamefont {Dai}, \citenamefont
  {Fang},\ and\ \citenamefont {Zhang}}]{zhang2009b}%
  \BibitemOpen
  \bibfield  {author} {\bibinfo {author} {\bibfnamefont {H.}~\bibnamefont
  {Zhang}}, \bibinfo {author} {\bibfnamefont {C.-X.}\ \bibnamefont {Liu}},
  \bibinfo {author} {\bibfnamefont {X.~L.}\ \bibnamefont {Qi}}, \bibinfo
  {author} {\bibfnamefont {X.}~\bibnamefont {Dai}}, \bibinfo {author}
  {\bibfnamefont {Z.}~\bibnamefont {Fang}}, \ and\ \bibinfo {author}
  {\bibfnamefont {S.~C.}\ \bibnamefont {Zhang}},\ }\href {\doibase
  10.1038/nphys1270} {\bibfield  {journal} {\bibinfo  {journal} {Nat. Phys.}\
  }\textbf {\bibinfo {volume} {5}},\ \bibinfo {pages} {438} (\bibinfo {year}
  {2009})}\BibitemShut {NoStop}%
\bibitem [{\citenamefont {Xia}\ \emph {et~al.}(2009)\citenamefont {Xia},
  \citenamefont {Qian}, \citenamefont {Hsieh}, \citenamefont {Wray},
  \citenamefont {Pal}, \citenamefont {Lin}, \citenamefont {Bansil},
  \citenamefont {Grauer}, \citenamefont {Hor}, \citenamefont {Cava},\ and\
  \citenamefont {Hasan}}]{xia2009}%
  \BibitemOpen
  \bibfield  {author} {\bibinfo {author} {\bibfnamefont {Y.}~\bibnamefont
  {Xia}}, \bibinfo {author} {\bibfnamefont {D.}~\bibnamefont {Qian}}, \bibinfo
  {author} {\bibfnamefont {D.}~\bibnamefont {Hsieh}}, \bibinfo {author}
  {\bibfnamefont {L.}~\bibnamefont {Wray}}, \bibinfo {author} {\bibfnamefont
  {A.}~\bibnamefont {Pal}}, \bibinfo {author} {\bibfnamefont {H.}~\bibnamefont
  {Lin}}, \bibinfo {author} {\bibfnamefont {A.}~\bibnamefont {Bansil}},
  \bibinfo {author} {\bibfnamefont {D.}~\bibnamefont {Grauer}}, \bibinfo
  {author} {\bibfnamefont {Y.~S.}\ \bibnamefont {Hor}}, \bibinfo {author}
  {\bibfnamefont {R.~J.}\ \bibnamefont {Cava}}, \ and\ \bibinfo {author}
  {\bibfnamefont {M.~Z.}\ \bibnamefont {Hasan}},\ }\href {\doibase
  10.1038/nphys1274} {\bibfield  {journal} {\bibinfo  {journal} {Nat. Phys.}\
  }\textbf {\bibinfo {volume} {5}},\ \bibinfo {pages} {398} (\bibinfo {year}
  {2009})}\BibitemShut {NoStop}%
\bibitem [{\citenamefont {Chen}\ \emph {et~al.}(2009)\citenamefont {Chen},
  \citenamefont {Analytis}, \citenamefont {Chu}, \citenamefont {Liu},
  \citenamefont {Mo}, \citenamefont {Qi}, \citenamefont {Zhang}, \citenamefont
  {Lu}, \citenamefont {Dai}, \citenamefont {Fang}, \citenamefont {Zhang},
  \citenamefont {Fisher}, \citenamefont {Hussain},\ and\ \citenamefont
  {Shen}}]{chen2009a}%
  \BibitemOpen
  \bibfield  {author} {\bibinfo {author} {\bibfnamefont {Y.~L.}\ \bibnamefont
  {Chen}}, \bibinfo {author} {\bibfnamefont {J.~G.}\ \bibnamefont {Analytis}},
  \bibinfo {author} {\bibfnamefont {J.-H.}\ \bibnamefont {Chu}}, \bibinfo
  {author} {\bibfnamefont {Z.~K.}\ \bibnamefont {Liu}}, \bibinfo {author}
  {\bibfnamefont {S.-K.}\ \bibnamefont {Mo}}, \bibinfo {author} {\bibfnamefont
  {X.~L.}\ \bibnamefont {Qi}}, \bibinfo {author} {\bibfnamefont {H.~J.}\
  \bibnamefont {Zhang}}, \bibinfo {author} {\bibfnamefont {D.~H.}\ \bibnamefont
  {Lu}}, \bibinfo {author} {\bibfnamefont {X.}~\bibnamefont {Dai}}, \bibinfo
  {author} {\bibfnamefont {Z.}~\bibnamefont {Fang}}, \bibinfo {author}
  {\bibfnamefont {S.~C.}\ \bibnamefont {Zhang}}, \bibinfo {author}
  {\bibfnamefont {I.~R.}\ \bibnamefont {Fisher}}, \bibinfo {author}
  {\bibfnamefont {Z.}~\bibnamefont {Hussain}}, \ and\ \bibinfo {author}
  {\bibfnamefont {Z.-X.}\ \bibnamefont {Shen}},\ }\href {\doibase
  10.1126/science.1173034} {\bibfield  {journal} {\bibinfo  {journal}
  {Science}\ }\textbf {\bibinfo {volume} {325}},\ \bibinfo {pages} {178}
  (\bibinfo {year} {2009})}\BibitemShut {NoStop}%
\bibitem [{\citenamefont {{Castro Neto}}\ \emph {et~al.}(2009)\citenamefont
  {{Castro Neto}}, \citenamefont {Peres}, \citenamefont {Novoselov},\ and\
  \citenamefont {Geim}}]{neto2009}%
  \BibitemOpen
  \bibfield  {author} {\bibinfo {author} {\bibfnamefont {A.~H.}\ \bibnamefont
  {{Castro Neto}}}, \bibinfo {author} {\bibfnamefont {N.~M.~R.}\ \bibnamefont
  {Peres}}, \bibinfo {author} {\bibfnamefont {K.~S.}\ \bibnamefont
  {Novoselov}}, \ and\ \bibinfo {author} {\bibfnamefont {A.~K.}\ \bibnamefont
  {Geim}},\ }\href {\doibase 10.1103/RevModPhys.81.109} {\bibfield  {journal}
  {\bibinfo  {journal} {Rev. Mod. Phys.}\ }\textbf {\bibinfo {volume} {81}},\
  \bibinfo {pages} {109} (\bibinfo {year} {2009})}\BibitemShut {NoStop}%
\bibitem [{\citenamefont {Novoselov}\ \emph {et~al.}(2005)\citenamefont
  {Novoselov}, \citenamefont {Geim}, \citenamefont {Morozov}, \citenamefont
  {Jiang}, \citenamefont {Katsnelson}, \citenamefont {Grigorieva},
  \citenamefont {Dubonos},\ and\ \citenamefont {Firsov}}]{novoselov2005}%
  \BibitemOpen
  \bibfield  {author} {\bibinfo {author} {\bibfnamefont {K.~S.}\ \bibnamefont
  {Novoselov}}, \bibinfo {author} {\bibfnamefont {a.~K.}\ \bibnamefont {Geim}},
  \bibinfo {author} {\bibfnamefont {S.~V.}\ \bibnamefont {Morozov}}, \bibinfo
  {author} {\bibfnamefont {D.}~\bibnamefont {Jiang}}, \bibinfo {author}
  {\bibfnamefont {M.~I.}\ \bibnamefont {Katsnelson}}, \bibinfo {author}
  {\bibfnamefont {I.~V.}\ \bibnamefont {Grigorieva}}, \bibinfo {author}
  {\bibfnamefont {S.~V.}\ \bibnamefont {Dubonos}}, \ and\ \bibinfo {author}
  {\bibfnamefont {A.~A.}\ \bibnamefont {Firsov}},\ }\href {\doibase
  10.1038/nature04233} {\bibfield  {journal} {\bibinfo  {journal} {Nature}\
  }\textbf {\bibinfo {volume} {438}},\ \bibinfo {pages} {197} (\bibinfo {year}
  {2005})}\BibitemShut {NoStop}%
\bibitem [{\citenamefont {Zhang}\ \emph {et~al.}(2005)\citenamefont {Zhang},
  \citenamefont {Tan}, \citenamefont {Stormer},\ and\ \citenamefont
  {Kim}}]{zhang2005}%
  \BibitemOpen
  \bibfield  {author} {\bibinfo {author} {\bibfnamefont {Y.}~\bibnamefont
  {Zhang}}, \bibinfo {author} {\bibfnamefont {Y.-W.}\ \bibnamefont {Tan}},
  \bibinfo {author} {\bibfnamefont {H.~L.}\ \bibnamefont {Stormer}}, \ and\
  \bibinfo {author} {\bibfnamefont {P.}~\bibnamefont {Kim}},\ }\href {\doibase
  10.1038/nature04235} {\bibfield  {journal} {\bibinfo  {journal} {Nature}\
  }\textbf {\bibinfo {volume} {438}},\ \bibinfo {pages} {201} (\bibinfo {year}
  {2005})}\BibitemShut {NoStop}%
\bibitem [{\citenamefont {Zhang}\ \emph {et~al.}()\citenamefont {Zhang},
  \citenamefont {Li}, \citenamefont {Feng}, \citenamefont {Kane},\ and\
  \citenamefont {Mele}}]{Zhang2013}%
  \BibitemOpen
  \bibfield  {author} {\bibinfo {author} {\bibfnamefont {F.}~\bibnamefont
  {Zhang}}, \bibinfo {author} {\bibfnamefont {X.}~\bibnamefont {Li}}, \bibinfo
  {author} {\bibfnamefont {J.}~\bibnamefont {Feng}}, \bibinfo {author}
  {\bibfnamefont {C.~L.}\ \bibnamefont {Kane}}, \ and\ \bibinfo {author}
  {\bibfnamefont {E.~J.}\ \bibnamefont {Mele}},\ }\href
  {http://arxiv.org/abs/1309.7682} {\bibinfo  {journal} {arXiv:1309.7682}\
  }\BibitemShut {NoStop}%
\bibitem [{\citenamefont {Wu}\ \emph {et~al.}()\citenamefont {Wu},
  \citenamefont {Shan},\ and\ \citenamefont {Yan}}]{Wu2014}%
  \BibitemOpen
\bibfield  {journal} {  }\bibfield  {author} {\bibinfo {author} {\bibfnamefont
  {S.-c.}\ \bibnamefont {Wu}}, \bibinfo {author} {\bibfnamefont
  {G.}~\bibnamefont {Shan}}, \ and\ \bibinfo {author} {\bibfnamefont
  {B.}~\bibnamefont {Yan}},\ }\href {http://arxiv.org/abs/1405.4731} {\bibinfo
  {journal} {arXiv:1405.4731}\ }\BibitemShut {NoStop}%
\bibitem [{\citenamefont {Kane}\ and\ \citenamefont
  {Mele}(2005{\natexlab{b}})}]{kane2005a}%
  \BibitemOpen
\bibfield  {journal} {  }\bibfield  {author} {\bibinfo {author} {\bibfnamefont
  {C.~L.}\ \bibnamefont {Kane}}\ and\ \bibinfo {author} {\bibfnamefont {E.~J.}\
  \bibnamefont {Mele}},\ }\href {\doibase 10.1103/PhysRevLett.95.146802}
  {\bibfield  {journal} {\bibinfo  {journal} {Phys. Rev. Lett.}\ }\textbf
  {\bibinfo {volume} {95}},\ \bibinfo {pages} {146802} (\bibinfo {year}
  {2005}{\natexlab{b}})}\BibitemShut {NoStop}%
\bibitem [{\citenamefont {Min}\ \emph {et~al.}(2006)\citenamefont {Min},
  \citenamefont {Hill}, \citenamefont {Sinitsyn}, \citenamefont {Sahu},
  \citenamefont {Kleinman},\ and\ \citenamefont {MacDonald}}]{min2006}%
  \BibitemOpen
  \bibfield  {author} {\bibinfo {author} {\bibfnamefont {H.}~\bibnamefont
  {Min}}, \bibinfo {author} {\bibfnamefont {J.~E.}\ \bibnamefont {Hill}},
  \bibinfo {author} {\bibfnamefont {N.~A.}\ \bibnamefont {Sinitsyn}}, \bibinfo
  {author} {\bibfnamefont {B.~R.}\ \bibnamefont {Sahu}}, \bibinfo {author}
  {\bibfnamefont {L.}~\bibnamefont {Kleinman}}, \ and\ \bibinfo {author}
  {\bibfnamefont {A.~H.}\ \bibnamefont {MacDonald}},\ }\href {\doibase
  10.1103/PhysRevB.74.165310} {\bibfield  {journal} {\bibinfo  {journal} {Phys.
  Rev. B}\ }\textbf {\bibinfo {volume} {74}},\ \bibinfo {pages} {165310}
  (\bibinfo {year} {2006})}\BibitemShut {NoStop}%
\bibitem [{\citenamefont {Wu}\ \emph {et~al.}(2007)\citenamefont {Wu},
  \citenamefont {Bergman}, \citenamefont {Balents},\ and\ \citenamefont {{Das
  Sarma}}}]{wu2007}%
  \BibitemOpen
  \bibfield  {author} {\bibinfo {author} {\bibfnamefont {C.}~\bibnamefont
  {Wu}}, \bibinfo {author} {\bibfnamefont {D.}~\bibnamefont {Bergman}},
  \bibinfo {author} {\bibfnamefont {L.}~\bibnamefont {Balents}}, \ and\
  \bibinfo {author} {\bibfnamefont {S.}~\bibnamefont {{Das Sarma}}},\ }\href
  {\doibase 10.1103/PhysRevLett.99.070401} {\bibfield  {journal} {\bibinfo
  {journal} {Phys. Rev. Lett.}\ }\textbf {\bibinfo {volume} {99}},\ \bibinfo
  {pages} {070401} (\bibinfo {year} {2007})}\BibitemShut {NoStop}%
\bibitem [{\citenamefont {Wu}(2008{\natexlab{a}})}]{wu2008b}%
  \BibitemOpen
  \bibfield  {author} {\bibinfo {author} {\bibfnamefont {C.}~\bibnamefont
  {Wu}},\ }\href {\doibase 10.1103/PhysRevLett.100.200406} {\bibfield
  {journal} {\bibinfo  {journal} {Phys. Rev. Lett.}\ }\textbf {\bibinfo
  {volume} {100}},\ \bibinfo {pages} {200406} (\bibinfo {year}
  {2008}{\natexlab{a}})}\BibitemShut {NoStop}%
\bibitem [{\citenamefont {Wu}\ and\ \citenamefont {{Das
  Sarma}}(2008)}]{wu2008a}%
  \BibitemOpen
  \bibfield  {author} {\bibinfo {author} {\bibfnamefont {C.}~\bibnamefont
  {Wu}}\ and\ \bibinfo {author} {\bibfnamefont {S.}~\bibnamefont {{Das
  Sarma}}},\ }\href {\doibase 10.1103/PhysRevB.77.235107} {\bibfield  {journal}
  {\bibinfo  {journal} {Phys. Rev. B}\ }\textbf {\bibinfo {volume} {77}},\
  \bibinfo {pages} {235107} (\bibinfo {year} {2008})}\BibitemShut {NoStop}%
\bibitem [{\citenamefont {Wu}(2008{\natexlab{b}})}]{wu2008}%
  \BibitemOpen
  \bibfield  {author} {\bibinfo {author} {\bibfnamefont {C.}~\bibnamefont
  {Wu}},\ }\href {\doibase 10.1103/PhysRevLett.101.186807} {\bibfield
  {journal} {\bibinfo  {journal} {Phys. Rev. Lett.}\ }\textbf {\bibinfo
  {volume} {101}},\ \bibinfo {pages} {186807} (\bibinfo {year}
  {2008}{\natexlab{b}})}\BibitemShut {NoStop}%
\bibitem [{\citenamefont {Zhang}\ \emph {et~al.}(2010)\citenamefont {Zhang},
  \citenamefont {Hung},\ and\ \citenamefont {Wu}}]{zhang2010}%
  \BibitemOpen
  \bibfield  {author} {\bibinfo {author} {\bibfnamefont {S.}~\bibnamefont
  {Zhang}}, \bibinfo {author} {\bibfnamefont {H.-h.}\ \bibnamefont {Hung}}, \
  and\ \bibinfo {author} {\bibfnamefont {C.}~\bibnamefont {Wu}},\ }\href
  {\doibase 10.1103/PhysRevA.82.053618} {\bibfield  {journal} {\bibinfo
  {journal} {Phys. Rev. A}\ }\textbf {\bibinfo {volume} {82}},\ \bibinfo
  {pages} {053618} (\bibinfo {year} {2010})}\BibitemShut {NoStop}%
\bibitem [{\citenamefont {Lee}\ \emph {et~al.}(2010)\citenamefont {Lee},
  \citenamefont {Wu},\ and\ \citenamefont {{Das Sarma}}}]{lee2010}%
  \BibitemOpen
  \bibfield  {author} {\bibinfo {author} {\bibfnamefont {W.-C.}\ \bibnamefont
  {Lee}}, \bibinfo {author} {\bibfnamefont {C.}~\bibnamefont {Wu}}, \ and\
  \bibinfo {author} {\bibfnamefont {S.}~\bibnamefont {{Das Sarma}}},\ }\href
  {\doibase 10.1103/PhysRevA.82.053611} {\bibfield  {journal} {\bibinfo
  {journal} {Phys. Rev. A}\ }\textbf {\bibinfo {volume} {82}},\ \bibinfo
  {pages} {053611} (\bibinfo {year} {2010})}\BibitemShut {NoStop}%
\bibitem [{\citenamefont {Zhang}\ \emph {et~al.}(2011)\citenamefont {Zhang},
  \citenamefont {Hung}, \citenamefont {Zhang},\ and\ \citenamefont
  {Wu}}]{zhang2011a}%
  \BibitemOpen
  \bibfield  {author} {\bibinfo {author} {\bibfnamefont {M.}~\bibnamefont
  {Zhang}}, \bibinfo {author} {\bibfnamefont {H.-h.}\ \bibnamefont {Hung}},
  \bibinfo {author} {\bibfnamefont {C.}~\bibnamefont {Zhang}}, \ and\ \bibinfo
  {author} {\bibfnamefont {C.}~\bibnamefont {Wu}},\ }\href {\doibase
  10.1103/PhysRevA.83.023615} {\bibfield  {journal} {\bibinfo  {journal} {Phys.
  Rev. A}\ }\textbf {\bibinfo {volume} {83}},\ \bibinfo {pages} {023615}
  (\bibinfo {year} {2011})}\BibitemShut {NoStop}%
\bibitem [{\citenamefont {Jacqmin}\ \emph {et~al.}(2014)\citenamefont
  {Jacqmin}, \citenamefont {Carusotto}, \citenamefont {Sagnes}, \citenamefont
  {Abbarchi}, \citenamefont {Solnyshkov}, \citenamefont {Malpuech},
  \citenamefont {Galopin}, \citenamefont {Lema\^{\i}tre}, \citenamefont
  {Bloch},\ and\ \citenamefont {Amo}}]{Jacqmin2014}%
  \BibitemOpen
  \bibfield  {author} {\bibinfo {author} {\bibfnamefont {T.}~\bibnamefont
  {Jacqmin}}, \bibinfo {author} {\bibfnamefont {I.}~\bibnamefont {Carusotto}},
  \bibinfo {author} {\bibfnamefont {I.}~\bibnamefont {Sagnes}}, \bibinfo
  {author} {\bibfnamefont {M.}~\bibnamefont {Abbarchi}}, \bibinfo {author}
  {\bibfnamefont {D.}~\bibnamefont {Solnyshkov}}, \bibinfo {author}
  {\bibfnamefont {G.}~\bibnamefont {Malpuech}}, \bibinfo {author}
  {\bibfnamefont {E.}~\bibnamefont {Galopin}}, \bibinfo {author} {\bibfnamefont
  {a.}~\bibnamefont {Lema\^{\i}tre}}, \bibinfo {author} {\bibfnamefont
  {J.}~\bibnamefont {Bloch}}, \ and\ \bibinfo {author} {\bibfnamefont
  {a.}~\bibnamefont {Amo}},\ }\href {\doibase 10.1103/PhysRevLett.112.116402}
  {\bibfield  {journal} {\bibinfo  {journal} {Phys. Rev. Lett.}\ }\textbf
  {\bibinfo {volume} {112}},\ \bibinfo {pages} {116402} (\bibinfo {year}
  {2014})}\BibitemShut {NoStop}%
\bibitem [{\citenamefont {Gemelke}\ \emph {et~al.}()\citenamefont {Gemelke},
  \citenamefont {Sarajlic},\ and\ \citenamefont {Chu}}]{gemelke2010}%
  \BibitemOpen
  \bibfield  {author} {\bibinfo {author} {\bibfnamefont {N.}~\bibnamefont
  {Gemelke}}, \bibinfo {author} {\bibfnamefont {E.}~\bibnamefont {Sarajlic}}, \
  and\ \bibinfo {author} {\bibfnamefont {S.}~\bibnamefont {Chu}},\ }\href
  {http://arxiv.org/abs/1007.2677} {\bibinfo  {journal} {arXiv:1007.2677}\
  }\BibitemShut {NoStop}%
\bibitem [{\citenamefont {Xu}\ \emph {et~al.}(2013)\citenamefont {Xu},
  \citenamefont {Yan}, \citenamefont {Zhang}, \citenamefont {Wang},
  \citenamefont {Xu}, \citenamefont {Tang}, \citenamefont {Duan},\ and\
  \citenamefont {Zhang}}]{xu2013}%
  \BibitemOpen
\bibfield  {journal} {  }\bibfield  {author} {\bibinfo {author} {\bibfnamefont
  {Y.}~\bibnamefont {Xu}}, \bibinfo {author} {\bibfnamefont {B.}~\bibnamefont
  {Yan}}, \bibinfo {author} {\bibfnamefont {H.-J.}\ \bibnamefont {Zhang}},
  \bibinfo {author} {\bibfnamefont {J.}~\bibnamefont {Wang}}, \bibinfo {author}
  {\bibfnamefont {G.}~\bibnamefont {Xu}}, \bibinfo {author} {\bibfnamefont
  {P.}~\bibnamefont {Tang}}, \bibinfo {author} {\bibfnamefont {W.}~\bibnamefont
  {Duan}}, \ and\ \bibinfo {author} {\bibfnamefont {S.-C.}\ \bibnamefont
  {Zhang}},\ }\href {\doibase 10.1103/PhysRevLett.111.136804} {\bibfield
  {journal} {\bibinfo  {journal} {Phys. Rev. Lett.}\ }\textbf {\bibinfo
  {volume} {111}},\ \bibinfo {pages} {136804} (\bibinfo {year}
  {2013})}\BibitemShut {NoStop}%
\bibitem [{\citenamefont {Wang}\ \emph {et~al.}()\citenamefont {Wang},
  \citenamefont {Xu},\ and\ \citenamefont {Zhang}}]{wang2014}%
  \BibitemOpen
  \bibfield  {author} {\bibinfo {author} {\bibfnamefont {J.}~\bibnamefont
  {Wang}}, \bibinfo {author} {\bibfnamefont {Y.}~\bibnamefont {Xu}}, \ and\
  \bibinfo {author} {\bibfnamefont {S.-C.}\ \bibnamefont {Zhang}},\ }\href
  {http://arxiv.org/abs/1402.4433} {\bibinfo  {journal} {arXiv:1402.4433}\
  }\BibitemShut {NoStop}%
\bibitem [{\citenamefont {Si}\ \emph {et~al.}(2014)\citenamefont {Si},
  \citenamefont {Liu}, \citenamefont {Xu}, \citenamefont {Wu}, \citenamefont
  {Gu},\ and\ \citenamefont {Duan}}]{si2014a}%
  \BibitemOpen
\bibfield  {journal} {  }\bibfield  {author} {\bibinfo {author} {\bibfnamefont
  {C.}~\bibnamefont {Si}}, \bibinfo {author} {\bibfnamefont {J.}~\bibnamefont
  {Liu}}, \bibinfo {author} {\bibfnamefont {Y.}~\bibnamefont {Xu}}, \bibinfo
  {author} {\bibfnamefont {J.}~\bibnamefont {Wu}}, \bibinfo {author}
  {\bibfnamefont {B.-L.}\ \bibnamefont {Gu}}, \ and\ \bibinfo {author}
  {\bibfnamefont {W.}~\bibnamefont {Duan}},\ }\href {\doibase
  10.1103/PhysRevB.89.115429} {\bibfield  {journal} {\bibinfo  {journal} {Phys.
  Rev. B}\ }\textbf {\bibinfo {volume} {89}},\ \bibinfo {pages} {115429}
  (\bibinfo {year} {2014})}\BibitemShut {NoStop}%
\bibitem [{\citenamefont {Liu}\ \emph {et~al.}()\citenamefont {Liu},
  \citenamefont {Guan}, \citenamefont {Song}, \citenamefont {Yang},
  \citenamefont {Yang},\ and\ \citenamefont {Yao}}]{liu2014}%
  \BibitemOpen
  \bibfield  {author} {\bibinfo {author} {\bibfnamefont {C.-C.}\ \bibnamefont
  {Liu}}, \bibinfo {author} {\bibfnamefont {S.}~\bibnamefont {Guan}}, \bibinfo
  {author} {\bibfnamefont {Z.}~\bibnamefont {Song}}, \bibinfo {author}
  {\bibfnamefont {S.~A.}\ \bibnamefont {Yang}}, \bibinfo {author}
  {\bibfnamefont {J.}~\bibnamefont {Yang}}, \ and\ \bibinfo {author}
  {\bibfnamefont {Y.}~\bibnamefont {Yao}},\ }\href
  {http://arxiv.org/abs/1402.5817} {\bibinfo  {journal} {arXiv:1402.5817}\
  }\BibitemShut {NoStop}%
\bibitem [{\citenamefont {Song}\ \emph {et~al.}()\citenamefont {Song},
  \citenamefont {Liu}, \citenamefont {Yang}, \citenamefont {Han}, \citenamefont
  {Ye}, \citenamefont {Fu}, \citenamefont {Yang}, \citenamefont {Niu},
  \citenamefont {Lu},\ and\ \citenamefont {Yao}}]{song2014}%
  \BibitemOpen
\bibfield  {journal} {  }\bibfield  {author} {\bibinfo {author} {\bibfnamefont
  {Z.}~\bibnamefont {Song}}, \bibinfo {author} {\bibfnamefont {C.-C.}\
  \bibnamefont {Liu}}, \bibinfo {author} {\bibfnamefont {J.}~\bibnamefont
  {Yang}}, \bibinfo {author} {\bibfnamefont {J.}~\bibnamefont {Han}}, \bibinfo
  {author} {\bibfnamefont {M.}~\bibnamefont {Ye}}, \bibinfo {author}
  {\bibfnamefont {B.}~\bibnamefont {Fu}}, \bibinfo {author} {\bibfnamefont
  {Y.}~\bibnamefont {Yang}}, \bibinfo {author} {\bibfnamefont {Q.}~\bibnamefont
  {Niu}}, \bibinfo {author} {\bibfnamefont {J.}~\bibnamefont {Lu}}, \ and\
  \bibinfo {author} {\bibfnamefont {Y.}~\bibnamefont {Yao}},\ }\href
  {http://arxiv.org/abs/1402.2399} {\bibinfo  {journal} {arXiv:1402.2399}\
  }\BibitemShut {NoStop}%
\bibitem [{\citenamefont {Wang}\ \emph
  {et~al.}(2013{\natexlab{a}})\citenamefont {Wang}, \citenamefont {Liu},\ and\
  \citenamefont {Liu}}]{wang2013a}%
  \BibitemOpen
\bibfield  {journal} {  }\bibfield  {author} {\bibinfo {author} {\bibfnamefont
  {Z.~F.}\ \bibnamefont {Wang}}, \bibinfo {author} {\bibfnamefont
  {Z.}~\bibnamefont {Liu}}, \ and\ \bibinfo {author} {\bibfnamefont
  {F.}~\bibnamefont {Liu}},\ }\href {\doibase 10.1103/PhysRevLett.110.196801}
  {\bibfield  {journal} {\bibinfo  {journal} {Phys. Rev. Lett.}\ }\textbf
  {\bibinfo {volume} {110}},\ \bibinfo {pages} {196801} (\bibinfo {year}
  {2013}{\natexlab{a}})}\BibitemShut {NoStop}%
\bibitem [{\citenamefont {Wang}\ \emph
  {et~al.}(2013{\natexlab{b}})\citenamefont {Wang}, \citenamefont {Liu},\ and\
  \citenamefont {Liu}}]{wang2013}%
  \BibitemOpen
  \bibfield  {author} {\bibinfo {author} {\bibfnamefont {Z.~F.}\ \bibnamefont
  {Wang}}, \bibinfo {author} {\bibfnamefont {Z.}~\bibnamefont {Liu}}, \ and\
  \bibinfo {author} {\bibfnamefont {F.}~\bibnamefont {Liu}},\ }\href {\doibase
  10.1038/ncomms2451} {\bibfield  {journal} {\bibinfo  {journal} {Nat.
  Commun.}\ }\textbf {\bibinfo {volume} {4}},\ \bibinfo {pages} {1471}
  (\bibinfo {year} {2013}{\natexlab{b}})}\BibitemShut {NoStop}%
\bibitem [{\citenamefont {Liu}\ \emph {et~al.}(2013)\citenamefont {Liu},
  \citenamefont {Wang}, \citenamefont {Mei}, \citenamefont {Wu},\ and\
  \citenamefont {Liu}}]{liu2013}%
  \BibitemOpen
  \bibfield  {author} {\bibinfo {author} {\bibfnamefont {Z.}~\bibnamefont
  {Liu}}, \bibinfo {author} {\bibfnamefont {Z.-F.}\ \bibnamefont {Wang}},
  \bibinfo {author} {\bibfnamefont {J.-W.}\ \bibnamefont {Mei}}, \bibinfo
  {author} {\bibfnamefont {Y.-S.}\ \bibnamefont {Wu}}, \ and\ \bibinfo {author}
  {\bibfnamefont {F.}~\bibnamefont {Liu}},\ }\href {\doibase
  10.1103/PhysRevLett.110.106804} {\bibfield  {journal} {\bibinfo  {journal}
  {Phys. Rev. Lett.}\ }\textbf {\bibinfo {volume} {110}},\ \bibinfo {pages}
  {106804} (\bibinfo {year} {2013})}\BibitemShut {NoStop}%
\bibitem [{\citenamefont {Schnyder}\ \emph {et~al.}(2008)\citenamefont
  {Schnyder}, \citenamefont {Ryu}, \citenamefont {Furusaki},\ and\
  \citenamefont {Ludwig}}]{schnyder2008}%
  \BibitemOpen
  \bibfield  {author} {\bibinfo {author} {\bibfnamefont {A.~P.}\ \bibnamefont
  {Schnyder}}, \bibinfo {author} {\bibfnamefont {S.}~\bibnamefont {Ryu}},
  \bibinfo {author} {\bibfnamefont {A.}~\bibnamefont {Furusaki}}, \ and\
  \bibinfo {author} {\bibfnamefont {A.}~\bibnamefont {Ludwig}},\ }\href
  {\doibase 10.1103/PhysRevB.78.195125} {\bibfield  {journal} {\bibinfo
  {journal} {Phys. Rev. B}\ }\textbf {\bibinfo {volume} {78}},\ \bibinfo
  {pages} {195125} (\bibinfo {year} {2008})}\BibitemShut {NoStop}%
\bibitem [{\citenamefont {Liang}\ \emph {et~al.}(2013)\citenamefont {Liang},
  \citenamefont {Wu},\ and\ \citenamefont {Hu}}]{Liang2013a}%
  \BibitemOpen
  \bibfield  {author} {\bibinfo {author} {\bibfnamefont {Q.-F.}\ \bibnamefont
  {Liang}}, \bibinfo {author} {\bibfnamefont {L.-H.}\ \bibnamefont {Wu}}, \
  and\ \bibinfo {author} {\bibfnamefont {X.}~\bibnamefont {Hu}},\ }\href
  {http://iopscience.iop.org/1367-2630/15/6/063031
  http://stacks.iop.org/1367-2630/15/i=6/a=063031?key=crossref.d84d43b6b61c7cdb4c2e7b9526e8f65c}
  {\bibfield  {journal} {\bibinfo  {journal} {New J. Phys.}\ }\textbf {\bibinfo
  {volume} {15}},\ \bibinfo {pages} {063031} (\bibinfo {year}
  {2013})}\BibitemShut {NoStop}%
\end{thebibliography}
\end{document}